\documentclass[12pt, a4paper]{article}

\usepackage{array}
\usepackage{booktabs,siunitx}
\usepackage{pgf-pie}      % 画饼图
\usepackage{subcaption}   % 可选：并排小图
\usepackage{booktabs}     % 只是风格统一（可删）
\usepackage{multirow}
\usepackage{makecell}   % 可用 \makecell 手动换行
\usepackage{tabularx}
\usepackage{geometry} % 页面设置
\usepackage{setspace} % 行距设置
\usepackage[round,authoryear]{natbib}  % 选项可按期刊要求调整
\usepackage{authblk}

\usepackage{graphicx} % Required for inserting images
\usepackage[ruled,boxed,linesnumbered]{algorithm2e}
\usepackage{xcolor}

\usepackage{hyperref}   % 建议放在导言区最后加载
\usepackage{url}
\usepackage{tablefootnote}

\usepackage{amsmath,amsthm,amssymb}
\usepackage{tikz}
\usetikzlibrary{arrows.meta, positioning,shapes.geometric,patterns}
\newcommand{\bfx}{\boldsymbol{x}}
\newcommand{\bfA}{\boldsymbol{A}}
\newcommand{\bfy}{\boldsymbol{y}}

\newcommand{\bff}{\boldsymbol{f}}

\newcommand{\bfb}{\boldsymbol{b}}
\newcommand{\bfa}{\boldsymbol{a}}
\newcommand{\bfJ}{\boldsymbol{J}}

\newcommand{\RR}{\mathbb{R}}

\newcommand{\tx}{\tilde{x}}
\newcommand{\IRR}{\mathbb{I}\RR}

\newcommand{\intbox}{% will this introduce white space?
{\,\,\setlength{\unitlength}{.33mm}\framebox(4,7){}\,}}

\DeclareMathOperator{\smear}{smear}
\DeclareMathOperator{\range}{range}
\DeclareMathOperator{\wid}{wid}
\DeclareMathOperator{\midd}{mid}

\newcommand{\cored}[1]{\textcolor{red}{#1}}
\newcommand{\ignore}[1]{}

\begin{document}

%\begin{frontmatter}



\title{A Dataset of Nonlinear Equations for Subdivision} %% Article title

%\title{A Survey and a Dataset for Subdivision} %% Article title

%use optional labels to link authors explicitly to addresses:

\author[1,2]{Juan Xu}
\author[1,2]{Huilong Lai}
\author[1,2]{Yingying Cheng}
\author[1,2]{Wenqiang Yang}
\author[1,2]{Changbo Chen\thanks{Corresponding author. Email: chenchangbo@cigit.ac.cn}}

\affil[1]{Chongqing Institute of Green and Intelligent Technology, Chinese Academy of Sciences, Chongqing, China}
\affil[2]{Chongqing School, University of Chinese Academy of Sciences, Chongqing, China}

%\affiliation[CIGIT]{organization={Chongqing Institute of Green and Intelligent Technology, Chinese Academy of Sciences},
            %addressline={},
%            city={Beibei District},
            %postcode={},
%            state={Chongqing},
%            country={China}}
%\affiliation[UCAS]{organization={Chongqing School, University of Chinese Academy of Sciences},
%            city={Beibei District},
            %postcode={},
%           state={Chongqing},
%            country={China},
%            }
% \author{} %% Author name

% \affiliation{organization={},%Department and Organization

%% Abstract

% transcendental systems

\maketitle
\begin{abstract}

\textbf{Purpose}: Subdivision is a basic and important  reliable method for solving systems of nonlinear equations, including both polynomial and transcendental equations. This work aims to provide a large real dataset for benchmarking and optimizing subdivision methods as well as  helping to develop machine learning methods to solve nonlinear equations. 
		
\textbf{Design\slash methodology\slash approach}\textbf{:} 
To build the dataset, we have searched the existing main benchmark suits for subdivision methods. 
We have also browsed over 1000 literatures and searched in detail over 300 literatures related to subdivision to find examples not covered by these repositories. After carefully comparing these systems and removing duplicate examples, we are left with a collection of 451 polynomial systems and 130 non-polynomial systems from the literatures and public datasets. To enhance the dataset, we further  generate 48000 zero-dimensional instances from  5 families of parametric nonlinear systems arising in practical applications. We leverage two subdivision solvers IbexSolve and RealPaver as well as a symbolic solver RootFinding:-Isolate in Maple to provide reliable solutions for these examples.

	\textbf{Findings}: 
% 	Key discoveries and results. analysis of the bug, analysis of some abnormal examples, reavealing the weakness, limitation of .
% 	Is robust on most the examples (most them is consistent with symbolic, for some special, we investigated and found an explanation). situations different from common sense. find cases where further optimization is needed. 
% 	Find opportunies to further optimize subdivision. It is effective on high-dimensional examples (why?)
By providing solution information for the examples in our dataset, we have run a large scale test 
on subdivision solvers and have the following key findings: There is a large overlap among the examples collected from the literatures/repositories; Overall, the two state-of-the-art subdivision solvers IbexSolve and RealPaver are more efficient 
than the symbolic solver RootFinding:-Isolate and IbexSolve is more efficient than RealPaver, however, 
none of them is universally faster than the others; Although the subdivision solvers are recognized to be reliable, 
our experiments suggest that in rare cases IbexSolve may miss out some solutions.

% IbexSolve is more efficient than RealPaver and RootFinding; 
% Although IbexSolve and RealPaver are well-known reliable subdivision solvers, ; 
% However, we also found its weakness on isolating real roots when ...; 
% Some bugs of IbexSolve were also caught. 
% Among ? total number of examples in these repositories, there are only ? number of 
% different examples. 
% Among ? examples appearing in ? articles we investigated, there are only ? mutually different new examples not in the existing repositories. 
% ? bugs were caught for IbexSolve. 
% Subdivision can be efficient on very high dimensional systems due to ...
% Subdivision is reliable on most of the examples it an solve. 
% When it is not reliable, the following cases can happen: 
% We also find some abnormal cases when using subdivision solvers, such as...
% By analyzing them, we find that....
% These findings may help to design more efficient strategies for subdivision. 

%Restrictions or potential weaknesses in the study.		
	\textbf{Research limitations}: 	The studies have been focusing on collecting zero-dimensional systems for subdivision.
	It does not cover literatures or repositories on using subdivision to solve positive dimensional systems, such as approximating curve or surface. Although some  literatures and repositories on  solving zero-dimensional systems using symbolic or homotopy methods are also included, the study primarily focuses on subdivision methods and does not systematically collect examples for non-subdivision methods.

	\textbf{Practical implications}:
	%The potential applications or practical benefits of the research.
	This dataset may be employed to benchmark different subdivision solvers. As the dataset contains the solution information obtained with the default strategies of two state-of-the-art subdivision solvers, it provides a solid baseline when developing new subdivision methods or inventing alternative solving strategies for existing subdivision methods. This labelled dataset could also be used to aid the development of machine learning techniques for subdivision methods. 
		
	\textbf{Originality/value}: 
	%The unique contribution and significance of the study.
	We provide so far the largest real dataset of zero-dimensional nonlinear systems suitable for solving with subdivision.  
	Great efforts have been taken to remove duplicate systems coming from different sources and to obtain reliable solution information of the systems in the dataset. This labelled dataset can potentially be useful for training and evaluating machine learning models in addition to serving as a benchmark for subdivision solvers. Another value of this work is to compare subdivision methods with symbolic methods, which was seldom systematically done before and the comparision sheds insights on the strength and weakness of both methods.
    %The dataset includes solution information for each collected system. 

    % number of literatures, time spent, compare the subdivision methods with symbolic methods, which come from diffenrent community, bug found due to large scale test. 

% \textbf{Purpose}: This study explores the ...
		
% 	\textbf{Design\slash methodology\slash approach}\textbf{:} The methods used in conducting the study.
		
% 	\textbf{Findings}: Key discoveries and results.
		
% 	\textbf{Research limitations}: Restrictions or potential weaknesses in the study.
		
% 	\textbf{Practical implications}:The potential applications or practical benefits of the research.
		
% 	\textbf{Originality/value}: The unique contribution and significance of the study.

%  Subdivision methods are a class of reliable methods for solving systems of nonlinear equations, including but not restricted to polynomial equations.
%     In this paper, we present a dataset of zero-dimensional square systems of nonlinear equations which are suitable for subdivision methods.
%     This database contains 453 polynomial systems and 130 {\color{red}non-polynomial systems} collected from the literature and public datasets, and 45000 nonlinear {\color{red} instances} generated from {\color{red} 5 families of parametric systems arising in} practical applications. 
%     Based on this dataset, we compared the performance of two existing popular software tools: IbexSolve and RealPaver. It turns out that IbexSolve is more efficient. 
%     By analyzing the strategies of 
%     We also compare the performance of the subdivision tools and the symbolic tools in Maple.
\end{abstract}


% \begin{highlights}
% \end{highlights}

%% Keywords
% \begin{keyword}
% Subdivision \sep 
% Nonlinear system solving \sep
% Dataset
% %% keywords here, in the form: keyword \sep keyword

% %% PACS codes here, in the form: \PACS code \sep code

% %% MSC codes here, in the form: \MSC code \sep code
% %% or \MSC[2008] code \sep code (2000 is the default)

% \end{keyword}

%\end{frontmatter}

% \documentclass{article}

% \section*{Todo}
% \begin{itemize}
%     \item Todo (before 2025-3-7)
%     \begin{itemize}
%         \item Run all the examples in Cheng's laptop and the server (Cheng).
%         \item Including running time/if all the solutions or not/number of solutions/solutions. (Cheng)
%         \item A .bib file containing all related literatures to subdivision. (Cheng)
%         \item Include the websites in the bib containing all the benchmark suites related to subdivision. (Cheng)
%         \item Theoretical/algorithmic Literatures on using subdivison for univariate/multivariate real root isolation. (Xu)
%         \item Relation of subdivision to exact polynomial system solving and literatures on parametric-system solving (Chen).  
%         \item Arguments and literatures on The unique value of subdivision (Exact solving/real root counting of non-polynomials). (Xu)
%         \item Cite other methods for computing solutions of non-poly sytems (Xu)
%         \item Combining subdivision with grid sampling for solving parametric-nonpoly systems (Chen). 
%         \item Start collecting most important/recent literatures (text book,classical,latest) on B\&B-optimization. (Cheng, Chen, Xu). 
%         \item A description of a complete subdivision algorithm (based on draft on another overleaf project and Julia implementation) (Xu). 
%         \item Try to identify several important classesd for subdivision (Cheng, Xu)
%     \end{itemize}
%     \item Next meeting (need to fix the important application classes)
% \end{itemize}

\section{Introduction}
  Solving systems of nonlinear equations is a fundamental task with application in various fields. Subdivision methods are a class of reliable methods for locating the real solutions of system of nonlinear equation within a bounded region, 
  the basic idea of which is to repeatedly partition the region of interest
  until the solutions are isolated or sufficiently approximated.
  Of course, subdivision is not limited to real-root computation; variants have been used for complex root isolation~\citep{becker2018near} and for global optimization~\citep{neumaier04,chabert2009contractor}. In this paper, however, we restrict attention to subdivision methods for computing real roots.
  Naturally, these methods fall under the Branch-and-Prune framework:
  regions proven to contain no solutions are pruned, those containing a unique solution (or sufficiently small) are retained, and uncertain regions are further branched into smaller regions.
  It can be observed that this framework shares many similarities with the Branch-and-Bound (B\&B) framework used for global optimization and is often discussed alongside it~\citep{neumaier04,chabert2009contractor}.

    Besides subdivision, symbolic methods~\citep{cox1997ideals,von2003modern} and homotopy continuation~\citep{sommese2005numerical} are 
    two principal approaches to solve nonlinear equations.
    Unlike subdivision, these approaches are mainly applicable to polynomial systems.
    Symbolic methods, or more precisely elimination methods, transform a polynomial system into triangular systems % not triangular sets
    which are easier to solve.
    Then real root isolation techniques, which are typically based on subdivision, can be applied to locate the solution of the system.
    Symbolic methods are theoretically elegant and rigorous, but are often costly in memory and computation because they rely on exact arithmetic.
    Homotopy continuation, by contract, solves a polynomial system by tracking solution paths from a start system with known (or easily computed) solutions to the target system. It is generally efficient but not inherently certified, as path jumping during tracking can lead to missed solutions.

    In many applications, interest is limited to solutions within a prescribed domain. In such cases, as noted in \citet{sherbrooke1993computation}, symbolic and homotopy methods often deliver more information than needed: symbolic elimination still transforms the system to triangular form, and homotopy continuation still tracks all complex solution paths; and these steps typically dominate the runtime. By contrast, subdivision restricts computation to the specified region, so the smaller the region, the lower the cost. 
    A further distinction is that: many subdivision methods apply to general nonlinear systems (including those with elementary functions), whereas symbolic and homotopy approaches are largely confined to polynomial systems.

  The Branch-and-Prune framework used in subdivision methods is closely related to the Branch-and-Bound framework.
  And heuristics have been playing an important role in Branch-and-Bound algorithms. 
  A successful application of these heuristics requires a balance between effectiveness and efficiency. 
  For instance, the strong branching heuristic has been shown be to very effective in reducing the size of the B\&B tree. 
  On the other hand, the cost of applying such a heuristic is high due to the need to solve linear programming problems 
  for all branching variables. 
  To reduce such a cost, one can apply supervised machine learning to imitate the behavior of strong branching. 
 Solely relying on imitation can hardly find heuristics more effective than those designed by humans. 
 Another important direction is to apply reinforcement learning to find solving strategies in a more autonomous way, aiming 
 to beat the best one invented by humans. 
 Although important progress has been made on integrating machine learning 
 to branch and bound algorithms~\citep{Scavuzzo2024}, a recent evaluation of ML-based solvers for combinatorial optimization, 
 including mixed integer linear programming, shows that 
 it is still difficult for learning based approaches to beat the best solvers~\citep{feng2025evaluationCO}, in particular on hard benchmark examples.
 Nevertheless, for both training and evaluating learning based approaches, a collection of high quality benchmark examples 
 are mandatory. 
 We are also interested in applying machine learning to accelerate subdivision. 
  To do this, a necessary step is to have a real dataset to assess different methods or heuristics.
 % For this reason, we propose to create a large dataset. 
  %Similar effort has been done for the ordering selection problem for CAD~\cite{?}. 

    In this paper, we report on the largest labelled dataset constructed so far for solving zero-dimensional nonlinear systems with subdivision-based methods. 
    A brief, non-exhaustive survey with emphasis on the literature from the past two decades is also provided to accompany with the dataset.
    The value of the dataset has been demonstrated through benchmarking existing implementations as well as being used 
    for learning to classify the real roots of parametric systems. To summarize, our contribution is three-fold:
    \begin{itemize}
        \item We provide a survey on the development of subdivision algorithms for solving zero-dimensional systems,  which may serve as a
        guide for understanding the dataset and a tutorial for beginners interested on optimizing subdivision. 
        % Special attention is paid on different heuristic strategies. We also propose a framework for learning with these strategies.
        
        \item We collect zero-dimensional examples for subdivision from a broad range of literatures and benchmark datasets on subdivision. 
              We run several different solvers on these examples and collect timing and solution information to form a labelled real dataset for subdivision. 
        \item We demonstrate the usefulness of this dataset through benchmarking different solvers, finding the potential room to improve subdivision methods, and training machine learning models to classify the number of real solutions of parametric nonlinear systems.
    \end{itemize}
\section{Notations and preliminaries}
This section provides an overview of the notation and preliminary background that will be used throughout the paper. Following the standardized notation proposed in \citet{kearfott2010standardized}, we use lowercase letters to denote scalars and vectors, 
boldface lowercase letters to denote intervals and boxes (interval vectors),
and uppercase letters to denote matrices. 
For notational simplicity, vectors are treated as column vectors by default, and transposes are omitted when clear from the context.

A (real, closed, nonempty) interval represented by $[a,b]$ with $a\le b$  is the set $\{x\in \RR\mid a\le x\le b\}$.
We also allow unbounded intervals by permitting endpoints in $\RR\cup\{-\infty,+\infty\}$. Let $\mathbb{IR}$ denote the set of all (possibly unbounded) real intervals.
A box can be considered as an interval vector $\bfx = (\bfx_1, \ldots , \bfx_n) \subseteq\mathbb{IR}^n$, which consists of all points $x=(x_1,\cdots,x_n)$ such that $x_k\in\bfx_k$ for any $ k=1,\ldots,n$.
A one-dimensional interval can be considered as a special case of a box.
For a box $\bfx$ with dimension greater than 1, we denote its  $i$-th component by $\bfx_i$.

Given an interval $\bfx$, denote by $\underline{\bfx}$ and $\overline{\bfx}$ the lower and upper bounds (endpoints) of $\bfx$, respectively.
The width and center of an interval $\bfx = [\underline{\bfx},\overline{\bfx}]$ are defined as 
$\wid(\bfx) = \overline{\bfx}-\underline{\bfx}$
and $\midd(\bfx) = \frac{\overline{\bfx}+\underline{\bfx}}{2}$, respectively.
The width of a box $\bfx= (\bfx_1, \ldots , \bfx_n)$ is defined as the the vector
$\wid(\bfx) = (\wid(\bfx_1), \ldots, \wid(\bfx_n))$ and the maximum width is then 
defined as 
$\max\wid(\bfx)=\max_{1\le i\le n}\wid(\bfx_i)$.
Similarly, the center of a box is defined as $\midd(\bfx) = (\midd(\bfx_1),\ldots,\midd(\bfx_n))$, and the center of an interval matrix is defined analogously, in a componentwise manner.
Unless stated otherwise, $\wid(\cdot)$ and $\midd(\cdot)$ are defined for bounded intervals (and boxes).
Denote by ${\rm int}(\bfx)$ the interior of the box $\bfx$.
Given a set $S\subset \RR$, denote by $\intbox S$ the smallest interval that contains $S$.

For any operation $\circ \in \{+, -, *\}$, the operation between two intervals $\boldsymbol{a}$ and $\boldsymbol{b}$ is defined by
\[
\boldsymbol{a} \circ \boldsymbol{b} = \{a \circ b \mid a \in \boldsymbol{a},\, b \in \boldsymbol{b}\}.
\]
Division and some basic elementary functions (without composition) between intervals can be defined analogously, although particular caution is needed if the divisor includes zero or if the function is not defined throughout the interval. Detailed formulas can be found in classic references such as~\citet{hansen2003global}.

In interval arithmetic, it is crucial to clearly distinguish between functions and expressions. While a function represents an abstract mathematical mapping from inputs to outputs, an expression refers to a specific syntactic arrangement of operations used to evaluate that function. Different expressions representing the same function can lead to significantly different interval evaluations due to the so-called dependency problem.
Conceptually we keep this distinction, but for notational convenience we shall usually use the same symbol, say $f$, both for a function and for one fixed expression representing it; the intended meaning will be clear from the context.
We use lowercase letters for scalar-valued functions and uppercase letters for vector-valued functions or expressions. Boldface letters are used to denote interval- or box-valued functions or expressions.

Given an expression $f$ and a box $\bfx$, the interval evaluation obtained by substituting the component intervals of $\bfx$ into the expression and performing the interval arithmetic operations is denoted as $\bff(\bfx)$.
Note that $\bff(\bfx)$ is typically not equal to the exact range of the function $f$ over the box $\bfx$, which is defined as $\range(f,\bfx) := \{f(x) \mid x \in \bfx\}$. Instead, due to the dependency problem, the exact range is always contained within the computed interval evaluation: $\range(f,\bfx) \subseteq \bff(\bfx)$.
The interval function $\bff(\bfx): \mathbb{IR}^n \to \mathbb{IR}$, constructed in this manner, is referred to as the \emph{natural interval  extension} of the function represented by $f$.
This extension provides an enclosure of the true range of $f$, and its inclusion property guarantees that the computed interval always contains the exact range.
In addition, the natural interval extension converges to the exact range of the function as the width of the input box tends to zero (assuming exact arithmetic and no rounding errors).
As a function may admit multiple algebraically equivalent expressions, it can also have multiple corresponding interval extensions. Different interval extensions can lead to different enclosures of the range.
 A notable example is the Horner extension, which rewrites a polynomial into a nested (Horner) form and then applies the natural interval extension to this reformulated expression.
This extension typically requires fewer interval operations and reduces the dependency effect, often resulting in tighter enclosures.
Another commonly used interval extension is the mean value extension.
Given a differentiable function 
$f$ and a box $\bfx$, the mean value extension of $f$ is constructed using the formula:
$$\bff_{\rm MV}(\bfx) = f(m) + \bfJ_f(\bfx)(\bfx-m)$$
where $m$ is the center of $\bfx$ and $\bfJ_f(\bfx)$ is an interval extension of the Jacobian matrix of $f$ over $\bfx$.

\section{A survey of subdivision}
    % {\color{red}An introduction sentence is needed.}
In this section, we present a survey of subdivision methods for certified real-root solving.
Unlike elimination and homotopy methods, which solve the original system by relating it to a more easily solvable one, either through algebraic transformation or continuous deformation, subdivision methods directly search for solutions of the original system within the prescribed domain of the variables, in a quite natural way.

We first present the general framework of subdivision methods, then discuss the key algorithmic ingredients, including verification operators and contractors. Finally, we review strategies for choosing and combining these ingredients.

\subsection{A framework (algorithm) for subdivision}\label{subsec:framework}
Subdivision algorithms typically follow this general framework: Given a box of interest $\bfx^{(0)}\subseteq \RR^n$, the algorithm first attempts to reduce the box as much as possible. Then it checks whether the box satisfies the output criterion.  If the criterion is met, the box is added to the output list; otherwise, the box is bisected, and the same procedure is recursively applied to the resulting sub-boxes.

% \begin{algorithm}[H]
% \caption{Framework}
% \label{alg:framework}
% \KwIn{A nonlinear system $F$ and an initial box of interest $B \subseteq \mathbb{R}^n$}
% \KwOut{List of boxes $\mathcal{L}$ satisfying the output criterion}
% \Begin{
% Set $\mathcal{W}=\{B\}$; $\mathcal{L}=\emptyset$\;
% \While{$\mathcal{W} \neq \emptyset$}{
%     Select and remove a box $B$ from $\mathcal{W}$ by a {\bf node selection strategy}\label{line:select}\;
%     Reduce $B$ by removing regions that are proven to contain no roots by a {\bf contraction strategy}\label{line:reduce} \; 
%     \If {$B$ is non-empty}{
%     Divide $B$ into a set of sub-boxes ${\cal B}$ by a {\bf partition strategy}\label{line:bisect}\; % a box B may be repeatedly processed
%     \eIf{$|{\cal B}|=1$ and ${\cal B}_1$ satisfies the output criterion by a {\bf check strategy}\label{line:output}}{
%         Add ${\cal B}_1$ to $\mathcal{L}$\;
%     }{
%         $\mathcal{W}=\mathcal{W}\cup {\cal B}$\;
%     }
% }
% }
% \Return{$\mathcal{L}$}
% }
% \end{algorithm}

    \begin{algorithm}[H]
    \caption{Framework}
    \label{alg:simple-framework}
    \KwIn{An initial box of interest $\bfx^{(0)} \subseteq \mathbb{R}^n$.}
    \KwOut{A list of boxes $\mathcal{L}$, each contained in $\bfx^{(0)}$ and satisfying the output criterion.}

    Set $\mathcal{W}=\{\bfx^{(0)}\}$; $\mathcal{L}=\emptyset$\;
    \While{$\mathcal{W} \neq \emptyset$}{
        Select and remove a box $\bfx^{(k)}$ from $\mathcal{W}$ by a {\bf node selection strategy}\label{line:select}\;
        Reduce $\bfx^{(k)}$ by removing regions that are proven to contain no roots by a {\bf contraction strategy}\label{line:reduce}\; 
        \If {$\bfx^{(k)}$ is non-empty}{
        \eIf{$\bfx^{(k)}$ satisfies the output criterion by a {\bf check strategy} \label{line:output}}{
            Add $\bfx^{(k)}$ to $\mathcal{L}$\;
        }{
            % Divide $B$ into a set of sub-boxes ${\cal B}$ by a {\bf partition strategy}\label{line:bisect}\; 
            Divide $\bfx^{(k)}$ into two sub-boxes  $\bfx^{(k,1)}, \bfx^{(k,2)}$  by a {\bf bisection strategy}\label{line:bisect}\; 
            Add $\bfx^{(k,1)}$ and $\bfx^{(k,2)}$ to $\mathcal{W}$\;
        }
        }
    }
    \Return{$\mathcal{L}$}\;
    \end{algorithm}
Alg.\ref{alg:simple-framework} outlines the general framework of the subdivision algorithms, whcih follows a branch-and-prune structure~\citep{caro2014branch}. 
%As a result, many algorithms developed under this framework are referred to in the literature as interval branch-and-prune algorithms~\cite{caro2014branch}. 
%Algorithms following this framework are often referred to as interval branch-and-prune algorithms, and they share many similarities with interval branch-and-bound algorithms used in global optimization.
%
The implementation of the above framework requires further specification and instantiation of a few key components: the node selection strategy in line~\ref{line:select}, the contraction strategy in line~\ref{line:reduce}, the bisection strategy in line~\ref{line:bisect}, and the check strategy in line~\ref{line:output}.
%The first two components will be discussed in detail in the next section.
%Now we briefly turn to the output criterion.
%
The choice of strategies is guided, on the one hand, by the objective of the algorithms and, on the other hand, by efficiency considerations.

The objective of the algorithm can be described from two perspectives: in terms of the number of outputs, it can be classified into three categories: finding a single solution, a specified number of solutions, or all solutions within a box; and in terms of the output requirement, it can be classified into two categories: root isolation or root approximation.

For root isolation, the check strategy is to detect whether the box $B$ contains exactly one root, typically achieved by a verification operator. For root approximation, the check strategy is to detect whether the size of $B$ falls below a prescribed tolerance $\varepsilon$.
In the case of root approximation, the algorithm is guaranteed to terminate, provided that the bisection strategy in line~\ref{line:bisect} is properly chosen. 
In contrast, for root isolation, it is generally difficult to prove that a box in the working list will eventually be confirmed by the verification operator, even if the box does contain exactly one root.
As a result, most algorithms in the literature introduce a tolerance as a fallback termination condition, and sufficiently small boxes are accepted as output, even though they may not be truly isolating.
As an exception, \citep{stahl1995thesis,xu2019} analyze the success conditions of the verification operator in the presence of a simple root, enabling termination without relying on such a tolerance.
It is important to note that verification operators are unable to confirm the existence of multiple roots. This limitation is a common challenge for numerical algorithms. In the presence of multiple roots, a tolerance is typically introduced to ensure the termination of the algorithm.

Most work in the literature aims to find all the roots in a given box. In this case, the node selection strategy does not affect either the output or the efficiency, so one can choose any strategy arbitrarily~\citep{araya2016interval}. A depth-first strategy is often preferred since it allows a box to be fully processed without switching regions. In contrast, when the goal is to find only one or a few roots, the node selection strategy becomes critical, as certain roots are easier to locate while others are more challenging, especially if they are ill-conditioned. Such cases are rare in the literature, with only a few exceptions.
In~\citet{huang2019evaluating}, where the goal is to find a single root, a depth-first selection strategy is still adopted; instead, the discussion is placed on the choice of the partitioning strategy.
In~\citet{bublitz2023thesis}, to quickly find a root of the system, the node selection strategy prioritizes the box at the center of which the sum of the absolute values of the function evaluations is minimal. Meanwhile, a heuristic root-finding procedure is also incorporated to accelerate the convergence to a root.
While node selection strategies are seldom investigated for zero-dimensional systems, they play an important role and have been discussed in Branch-and-Prune algorithms for positive-dimensional systems~\citep{delatour2021strategies, chenouard2009search}, and especially in Branch-and-Bound algorithms for global optimization~\citep{neveu2016node}.

In any context, both the contraction strategy and the bisection strategy are critical for efficiency of algorithms. We will discuss these aspects in section~\ref{subsec:strategy}, after introducing the contractors.

\subsection{Verification operators}
%The branching operator splits a region of interests into at least two parts, in the hope that the the solutions in smaller regions will be easier to identify. 
%Typically, the branching operator split the region of interests, usually represented as a box, into two sub-boxes, by bisecting along a certain dimension.
%The reduction operator removes the region that can be determined to contain no solutions. 
%It returns one or more boxes that are contained in the input box.
%This operator can be implemented by using consistency techniques such as box consistency~\cite{van1997solving}  or hull consistency, or interval Newton type techniques such as Krawczyk method or Hansen-Sengupta method.

%In practice, it may happen that two solutions of a system are very close to each other so that boxes with width greater than machine precision will not be able to separate them.
%In such case, one may need to use multiple precision.
%However, it may even happen that the system has a multiple solution, in which case the subdivision algorithm will not be able to terminate since the verification operator will keep failing.
%The inability to tackle multiple solutions is a common issue for numerical algorithms.
%To prevent the termination issue caused by the presence of multiple solutions and to improve the stability of the algorithm, one usually sets a  tolerance $\varepsilon$,  and stops to process the boxes in $Q$ when their size are smaller than $\varepsilon$.
%Such boxes with be moved to a list of suspect boxes.

The verification operator checks whether there is exactly one solution inside a box. 
In subdivision algorithms, the most commonly used verification operators are interval Newton-type operators such as Krawczyk operator and Hansen-Sengupta operator.

The Krawczyk operator, derived from fixed-point theorem, is particularly popular due to its simplicity and ease of implementation. 
Given a box $\bfx$, the Krawczyk operator is defined as
$$K(\bfx,\tx):=\tx-C\bff(\tx)+(\mathbb{I}-C\cdot \bfJ_f(\bfx))(\bfx-\tx)$$
where $\tx$ is a point in $\bfx$, $C$ is a preconditioning matrix, $\bff$ and $\bfJ_f$ are respectively interval extensions of $f$ and the Jacobian of $f$.
It has been proven that the Krawczyk operator yields the best (tightest) results when $\tx = \midd(\bfx)$ and $C=\midd(\bfJ_f(\bfx))^{-1}$.
The Krawczyk operator $K(\bfx,\tx)$ has the following properties~\citep{neumaier04}:
\begin{enumerate}
    \item Any zero $x^*\in \bfx$ of $f$ lies in $\bfx\cap K(\bfx,\tx)$.
    \item If $\bfx\cap K(\bfx,\tx)=\emptyset$, then $\bfx$ contains no zero of $f$.
    \item If $K(\bfx,\tx)\subseteq {\rm int}(\bfx)$ then $\bfx$ contains a unique zero of $f$.
\end{enumerate}
These properties show that the Krawczyk operator can serve both as a verification operator and as a contractor.
While the Krawczyk operator is simple and widely used in practice, it is not as powerful as Hansen-Sengupta operator~\citep{neumaier90}.
To introduce the definition of Hansen-Sengupta operator, we define a function for intervals $\bfa, \bfb, \bfx \in \IRR$:
        $$
        \Gamma(\bfa, \bfb, \bfx) := \intbox \big\{ 
        x \mid ax=b \text{ for some } a\in\bfa, b\in \bfb
        \big\}.
        $$
%
% In some definitions, \bfx is restricted to be within the box B,
% but in this case, one may need to reinterpret in the properties,
% otherwise the last property will always hold, which is not what we want.
%
    The function can be extended to interval matrix and vectors as 
        $
        \Gamma(\bfA, \bfb, \bfx) := \bfy
        $
    where $\bfA\in\IRR^{n\times n}, \bfb\in \IRR^n, \bfx\in\IRR^n$, and the $i$-th component $\bfy_i$ of $\bfy$ is defined as
        \begin{equation}\label{eq:hensen-iteration}
        \bfy_i = \Gamma(
            \bfA_{ii}, \bfb_i - \sum_{j<i} \bfA_{ij} \bfy_i 
                            - \sum_{j>i} \bfA_{ij} \bfx_i, 
            \bfx_i                
        ), \quad  i=1,\ldots, n.
       \end{equation}
    The operator $\Gamma(\bfA, \bfb, \bfx)$ is known as the interval Gauss-Seidel operator, which is used for solving interval linear systems.
    It provide an enclosure of the following set
    \begin{equation}
         \big\{
        x \mid Ax=b \text{ for some } A\in \bfA, b\in \bfb 
            \big\}.       
    \end{equation}
    Given a box $\bfx$, the Hansen-Sengupta operator is defined as
    \begin{equation}\label{eq:hansen}
        H(\bfx,\tx) := \tx + \Gamma(C\cdot\bfJ_f(\bfx), -C\bff(\tx),
        \bfx- \tx)   
    \end{equation}
    where $\tx$ is a point in $\bfx$ and $C$ is a preconditioning matrix.
    This operator shares the same nice properties as the Krawczyk operator described earlier.
    In addition, when used as a contractor, the Hansen–Sengupta operator is often more effective in practice, as it consistently produces tighter, or at least equally tight, enclosures for the solution set.

    Another variant of interval Newton-type operators can be obtained by replacing the interval Gauss-Seidel operator in~\eqref{eq:hansen} with the interval version of Gaussian elimination~\citep{neumaier04,neumaier90}.
    This new operator preserves the key properties of interval Newton-type methods and experiments show that it is more powerful than Hansen-Sengupta operator~\citep{jaulin2001applied}.
    However, due to its higher computational cost, it is less commonly used in practice.

    As mentioned above, interval-newton type can be used as a verification operator which confirms the uniqueness of a root within a box.
    In addition to these, there exist other tests, such as the Miranda test and the Borsuk test, that can verify only the existence, but not the uniqueness, of a solution within a box.
    The Miranda test confirms the existence of a zero of a continuous function in a box by checking for sign changes on opposite faces of the box. It can be viewed as an application of the intermediate value theorem in higher dimensions.
    The Borsuk test is a topological method for verifying the existence of a root by exploiting the antipodal properties of continuous functions over symmetric domains.
    It is theoretically powerful but more complex to implement, and thus rarely used in algorithms.
    It has been shown to be at least as powerful as the Miranda test~\citep{frommer2005existence}.
    In practice, the Miranda test is typically applied after preconditioning~\citep{mk1980}.
    A standard implementation of the Miranda test is shown to be at least as powerful as the existence test based on the Krawczyk operator~\citep{frommer2004comparison}, but less powerful than the one based on the Hansen–Sengupta operator~\citep{Goldsztejn2007comparison}, with similar computational cost.

\subsection{Contractors}
    As described earlier, interval Newton-type operators can be used as contractors. However, they are computationally expensive, as they involve computations over the entire system, including evaluating the interval Jacobian and solving interval linear system for all variables simultaneously.
   When applied to large boxes in the early stages of the search, the entries of the interval Jacobian matrix tend to have large widths, which may lead to poor performance of the operators.
    In such cases, the reduction achieved may not justify the computational cost.
    Consistency techniques provide an alternative approach to box reduction. Unlike interval Newton-type methods, which operate on the entire system, consistency techniques act on individual constraints (i.e., equations). Each constraint is used independently to reduce the domains of the variables involved.
    Compared to interval Newton methods, consistency techniques incur lower computational cost, and are better suited for quickly pruning large boxes during the early stages of the search. % Juan: relaxation and 3B are consistency tech? they are cheaper than interval newton? need to rethink and rewrite.
    
    Here we describe some common consistency techniques used in continuous constraint satisfaction problems, among which hull consistency and box consistency are the most widely used. 
    These techniques are applied to a single constraint at a time. In general, a constraint may be an equation or an inequality; in this paper, we focus on equations, i.e., constraints of the form 
    $f(x)=0$.
%    These methods aim to reduce the domains of variables by analyzing constraints, thereby pruning the search space while preserving all potential solutions. Although both techniques are grounded in interval analysis, they differ in pruning strength, computational cost, and the way they process constraints.

    Hull consistency, also known as 2B-consistency, aims to compute the smallest box that encloses all solutions of a given equation.
    A box $\bfx = (\bfx_1, \cdots ,  \bfx_n)$ is hull consistent with respect to a constraint $c(x_1,\ldots,x_n)$ if and only if $\forall i =1,\ldots,n$
    $$\bfx_i = \intbox \{ r_i \in \bfx_i \mid \exists r_1\in \bfx_1,\ldots,r_{i-1}\in \bfx_{i-1}, r_{i+1} \in \bfx_{i+1},\ldots, r_n\in I_n : c(r_1,\ldots,r_n)\}.$$
    In other words, a box $\bfx$ is hull consistent if, in each dimension, the corresponding interval cannot be further reduced, because the projection of the solution set onto that dimension covers both endpoints.
    Hull consistency is typically enforced through 
    forward-backward propagation~\citep{benhamou1999revising}.

    Given a constraint, forward-backward propagation operates on the variable domains in two directions:
    Forward evaluation computes the interval image of the constraint function over current box of the input variables.
    Backward evaluation attempts to refine the interval of each variable involved by intersecting it with the pre-image that could produce the computed output interval. 
    
 \begin{figure}[htbp]
  \centering

  \tikzset{
    op/.style={draw, circle, minimum size=8mm},
    var/.style={draw, rectangle, minimum size=6mm},
    const/.style={draw, diamond, inner sep=2pt, minimum size=0mm},
    val/.style={font=\scriptsize},
    arrowf/.style={->, thick, >=Stealth, blue},
    arrowb/.style={->, thick, >=Stealth, red}
  }

  \begin{tikzpicture}[scale=0.8, every node/.style={transform shape}]

    % ---------- FORWARD tree (left side) ----------
    \begin{scope}[xshift=-4.5cm]
    \node[op] (eq) {=}
      child[xshift=-1cm]
      {
        node[op] (exp) {$\hat{}$}
          child {node[const] (c1) {2} }
          child {node[var] (x) {$x$}
            node[val,below=2pt of x] {[-5,5]}
          }        
        node[val,left=5pt of exp] {[$\frac{1}{32},32$]}
      }
      child[xshift=1cm]
      {node[op] (plus) {$+$}
        child {node[var] (z) {$z$}
          node[val,below=2pt of z] {[2,10]}
        }
        child {node[op] (pow) {$\hat{}$}
          child {node[var] (y) {$y$}
            node[val,below=2pt of y] {[-5,5]}
          }
          child {node[const] (c2) {2} }
          node[val,right=5pt of pow] {[0,25]}
        }
        node[val,right=5pt of plus] {[2,35]}
      };

    \draw[arrowf] (x) -- (exp);
    \draw[arrowf] (c1) -- (exp);
    \draw[arrowf] (exp) -- (eq);
    \draw[arrowf] (z) -- (plus);
    \draw[arrowf] (y) -- (pow);
    \draw[arrowf] (c2) -- (pow);
    \draw[arrowf] (pow) -- (plus);
    \draw[arrowf] (plus) -- (eq);
    \end{scope}

    % ---------- BACKWARD tree (right side) ----------
    \begin{scope}[xshift=4.5cm]
    \node[op] (eq2) {=}
      child[xshift=-1cm]
      {
        node[op] (exp2) {$\hat{}$}
          child {node[const] (c12) {2} }
          child {node[var] (xb) {$x$}
            node[val,below=2pt of xb] {[\cored{1},5]}
          }
        node[val,left=5pt of exp2] {[\cored{2},32]}
      }
      child[xshift=1cm]
      {
        node[op] (plus2) {$+$}
          child {node[var] (zb) {$z$}
            node[val,below=2pt of zb] {[2,\cored{7}]}
          }
          child {node[op] (pow2) {$\hat{}$}
            child {node[var] (yb) {$y$}
              node[val,below=2pt of yb] {[-5,5]}
            }
            child {node[const] (c22) {2} }
            node[val,right=5pt of pow2] {[0,25]}
          }
          node[val,right=5pt of plus2] {[2,\cored{32}]}
      };

    \draw[arrowb] (eq2) -- (exp2);
    \draw[arrowb] (eq2) -- (plus2);
    \draw[arrowb] (exp2) -- (c12);
    \draw[arrowb] (exp2) -- (xb);
    \draw[arrowb] (plus2) -- (zb);
    \draw[arrowb] (plus2) -- (pow2);
    \draw[arrowb] (pow2) -- (yb);
    \draw[arrowb] (pow2) -- (c22);
    \end{scope}
  \end{tikzpicture}
  \caption{Forward (left) and backward (right) propagation for the constraint $2^x=z+y^2$.}
  \label{fig:forward-backward-tree}
\end{figure}
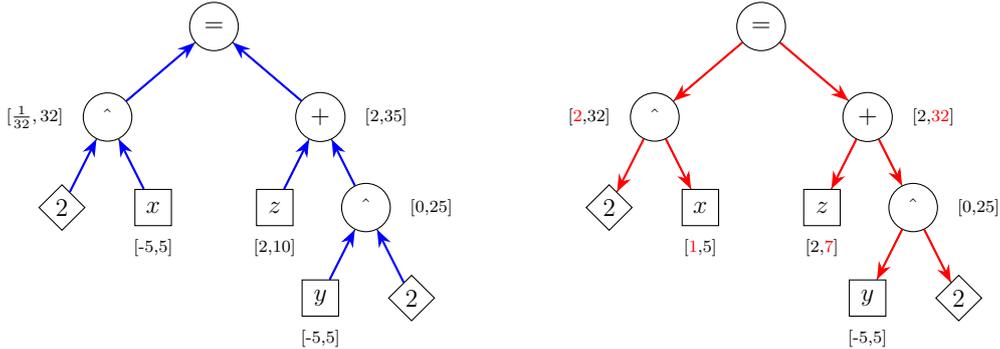

    Notice that forward-backward propagation attempts to reduce all the variables involved simultaneously. If each variable appears only once in the constraint, forward-backward propagation yields an optimal result, that is, the resulting box is hull consistent.
    However, when variables occur multiple times, each repeated occurrence is treated as an independent new variable during the evaluation process. 
    {This breaks the inherit dependency structure and may cause significant overestimation, thereby reducing the pruning efficiency of the method.}

    In case of multiple occurrence of variables, box consistency may provide better result.
    Unlike hull consistency, box consistency technique refines the interval of each variable independently by turning the original multivariate constraint into a univariate one, substituting all other variables with their interval.
    
    Consider a constraint 
    $f(x_1,\ldots, x_n) = 0$ and $\bff$ an interval extension of $f$.
     A box $(\bfx_1,\cdots , \bfx_n)\in \IRR^n$ is box consistent with respect to $\bff$ if and only if for $\forall i = 1,\ldots, n$,  the following holds:
    \[
    \left\{
    \begin{aligned}
    &0 \in \bff(\bfx_1,\ldots,\bfx_{i-1}, [\underline{\bfx_i}, \underline{\bfx_i}^+], \bfx_{i+1},\ldots,\bfx_n) \\
    &0 \in \bff(\bfx_1,\ldots,\bfx_{i-1}, [\overline{\bfx_i}^-, \overline{\bfx_i}], \bfx_{i+1},\ldots,\bfx_n)
    \end{aligned}
    \right.
    \]
    where $x^{+}$ (resp. $x^{-}$) denotes the floating-point successor of $x$ (resp. the floating-point predecessor of $x$).
    Observe that box consistency is defined with respect to the interval extension $\bff$ of $f$. For the same function $f$, different interval extensions may induce different forms of box consistency~\citep{van1997solving}.
    To enforce box consistency for reducing the box $(\bfx_1,\cdots , \bfx_n)$, we fix each variable $x_j$ for $j\neq i$ to their current interval $\bfx_j$ and consider the resulting interval function 
    $$\bff_i = \bff(\bfx_1,\ldots,\bfx_{i-1}, \bfx, \bfx_{i+1},\ldots,\bfx_n).$$
    This yields a univariate interval function of $\bfx$.
     To prune the interval $\bfx_i$, it amounts to
    tightening $\bfx_i$ until its endpoints are close to the leftmost and rightmost zeros of $\bff_i$.
    This can be achieved by recursively splitting the interval, pruning infeasible subintervals, and searching for the desired zeros within the leftmost and rightmost subintervals that have not yet been eliminated. The univariate interval Newton method may be used to improve the efficiency of this process. A detailed discussion on the efficient implementation of contractors based on box consistency is presented in~\citet{goldsztejn2010box}.
    
    Unlike hull consistency, which is defined using existential quantification, box consistency evaluates the interval extension of the constraint by plugging in the intervals of the other variables.
   {By definition, hull consistency appears to be stricter than box consistency.
    However, in practice, the box consistency technique produces tighter result than the forward-backward propagation when variables appear multiple times in the constraint.} 
    This is because box consistency focuses on one variable at a time and by searching for its leftmost and rightmost zeros, it effectively reduces the impact of variable dependency in that dimension~\citep{collavizza1999comparing}.
    On the other hand, hull consistency is less costly and has the potential to reduce the domains of all involved variables simultaneously. 

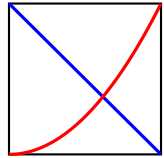
\begin{figure}[htbp]
    \centering
    \begin{tikzpicture}[scale=1] 

    \draw[thick] (-1,-1) rectangle (1,1);
    \draw[blue, very thick]
    plot[domain=-1:1] (\x, {-1*\x});
    \draw[red, very thick]
    plot[domain=-1:1, samples=200] (\x, {(\x+1)^2/2-1});
    \end{tikzpicture}
    \caption{The box $([-1,1],[-1,1])$ is hull consistent and box consistent with respect to the system $\{x_2=-x_1, 2x_2=(x_1+1)^2-2\}$.}
    \label{fig:stuck}
    \end{figure}

    Since hull consistency and box consistency are defined with respect to a single constraint, they are referred to as local consistency. 
    In a subdivision algorithm, it is necessary to propagate these consistency techniques across all constraints repeatedly until further reductions become negligible.
    However, since the constraints are treated individually, it may happen that the box fails to converge to the solution, even after all constraints have been processed, as shown in Fig.~\ref{fig:stuck}.
    In such case, stronger consistency techniques can be applied to further contract the box.

%    Interval-Newton-type contractors, as described above, can 
% 3B-consistency is also partial consistency
% check the definition of partial consistency and maybe rewrite this parag.
    A box $\bfx = (\bfx_1,\cdots , \bfx_n)$ is said to be 3B-consistent with respect to a set of constraints $\mathcal{S}$ if and only if for every $i =1,\ldots, n$,
    the box $\bfx$ is 2B-consistent (i.e. hull consistent) with respect to
    the reduced systems
    $$\mathcal{S}\cup \left\{x_i = \underline{\bfx_i}\right\}\quad
    \text{and}\quad 
    \mathcal{S}\cup \left\{x_i = \overline{\bfx_i}\right\}.$$
    It is easy to see that the box in Fig~\ref{fig:stuck} is not 3B-consistent for the constraints,
    since, for instance, the box is not 2B-consistent to the system $\{x_2=-x_1,2x_2=(x_1+1)^2-2, x_1=-1\}$.
    To reduce the box $\bfx$ in the $i$-th dimension by the definition of 3B-consistency, one can probe the interval $\bfx_i$ from both ends, and try to prove the infeasibility of small boundary subintervals using 2B-consistency techniques such as forward-backward propagation.
    For the example in Fig.~\ref{fig:stuck}, one can prove the infeasibility of the subbox $([-1,0], [-1,1])$ using 3B-consistency.
    3B-consistency technique is expected to be more powerful than box consistency and hull consistency techniques, while it is more expensive.
    If, in the definition of $3$B-consistency, the hull-consistency is replaced by box consistency, that is, $\bfx$ is  box consistent with respect to the reduced systems, then the resulting notion is called bound-consistency.
    Please refer to~\citet{collavizza1999comparing} for comparisons of the power of the consistency techniques described above.

    The definition of 3B-consistency can be generalized to $k$B-consistency by induction:
    A box $\bfx = (\bfx_1, \cdots ,  \bfx_n)$ is said to be $k$B-consistent with respect to a set of constraints $\mathcal{S}$ if and only if fo every $i =1,\ldots, n$,
    the box $B$ is $(k-1)$B-consistent with respect to 
    $$\mathcal{S}\cup \left\{x_i = \underline{\bfx_i}\right\}\quad
    \text{and}\quad 
    \mathcal{S}\cup \left\{x_i = \overline{\bfx_i}\right\}.$$
    However, $k$B technique for $k>3$ is rarely used in practice since it is too expensive.

    In~\citet{chabert2007cid}, the authors introduced two improved contractors, CID and 3BCID. Unlike standard 3B-consistency based contractors, which only attempt to remove boundary subboxes at the two ends of a chosen dimension, these methods also examine subboxes in the interior. They apply a hull-consistency contractor to these intermediate regions, thereby enabling additional reductions of the interior subboxes—and consequently of the whole box $\bfx$—in several, and potentially all, dimensions. Further improvements to 3BCID were later proposed in~\citet{neveu2015adaptive}.
    
    Another effective method for reducing the size of search boxes is linear relaxation. 
    The core idea is to enclose the solution set of each non-linear equation with a polytope defined by a set of linear inequalities,
    By solving the resulting linear problems, one can derive improved bounds in each dimension.
    Affine arithmetic offers an approach for constructing linear relaxation of the system.
    As an extension of interval arithmetic~\citep{kolev2001automatic,kolev2004improved,messine2002extentions},
 affine arithmetic preserves variable dependencies by representing each variable as an affine combination of noise symbols, resulting in tighter bounds than interval arithmetic.
    By applying affine arithmetic operations to all expressions in the constraints, these constraints are transformed into affine forms. The resulting affine representations naturally yield linear relaxations of the constraints~\citep{ninin2015reliable}.
    
    \citet{lebbah2005rigorous,lebbah2005efficient} proposed a linear relaxation technique for polynomial systems, adapted from reformulation-linearization technique.
    This method replaces nonlinear terms with auxiliary variables and introduces corresponding identity constraints into the original system.
    Linear inequalities are then constructed to approximate the semantics of these nonlinear term.
    By exploiting the specific structure of quadratic terms, this technique achieves tight enclosures of the solution set.
    Domes \& Neumaier~\citet{domes2010constraint} introduced an alternative relaxation method for quadratic constraint, which can be naturally extended to more general polynomial constraint. 
    The main difference between these two relaxation approach lies in the way that bilinear constraints are handled.
    A integration and comparison of these linear relaxation techniques, along with experimental evaluations, is presented in~\citet{domes2012rigorous}.
    
    More recently, \citet{araya2012contractor} introduced
    a linear relaxation technique based on interval Taylor expansions evaluated at a pair of opposite vertices of a box. 
    In~\citet{araya2025hybridizing}, the author proposes a hybrid approach that combines the affine relaxation technique with the extremal Taylor method, achieving better performance on constrained global optimization instances.

    As mentioned earlier, interval-Newton-type operators, such as the Krawczyk operator and Hansen-Sengupta operator, can also serve as contractors by virtue of their first property.
    When used as a contractor, the Krawczyk operator remains unchanged, whereas the Hansen-Sengupta operator requires a slight modification: in each iteration~(\ref{eq:hensen-iteration}), $\bfy_i$ is intersected with $\bfx_i$ to obtain a tighter enclosure.
    In addition, the interval Gauss-Seidel operator can be applied recursively until further improvement becomes negligible.
    It can be seen that the interval-Newton-type operators act on the entire system and have the potential to reduce the box in each direction.
    Compared to previously described contractors, which impose no restrictions on the shape of the system, interval-Newton-type methods require the system to be square.

%    \subsection{Satisfiability test operator}

\subsection{Strategy of solvers}~\label{subsec:strategy}
    The previous subsection introduced the key ingredients of the subdivision framework. The performance of a practical solver, however, depends critically on the strategies used to orchestrate these ingredients. We have already discussed about node-selection and checking strategies in Section~\ref{subsec:framework}. Here we focus on the contraction strategy and the bisection strategy.

    For contraction, one must decide in what order to apply contractors, how aggressively to iterate propagation, and when to switch from contraction to branching. For bisection, one must decide how to partition a box. While the bisection operation may appear straightforward, it can significantly affect the overall performance, and its impact on the efficiency is often harder to predict than that of the contraction strategy.

    It is often noted in the literature that bisection should be avoided as much as possible, especially when the number of variables is high~\citep{jaulin2001applied}. Typically, it is used only as a last resort, when the available contractors yield little or no significant reduction of the box, since a bisection in one direction is often followed by further bisections in others, resulting in exponential complexity.
    
    \subsubsection{Strategy on the contractors}
    As noted in~\citet{jaulin2001applied}, no contractor can be regarded as universally superior to the others. Each is designed to be effective under specific conditions.
    For instance, backward-forward propagation (based on hull consistency) is generally effective and has the potential to reduce a box along all dimensions. However, its effectiveness can be significantly diminished when a variable appears multiple times within a constraint.
    In contrast, contractors based on box consistency are more costly and reduce only one variable at a time. However, they can yield better results than forward-backward propagation.
    As mentioned above, these two type of contractors are local, and are well-suited for use in the early stages of the algorithm.
    However,  relying solely on them may lead to the situation of deadlock, where the box can no longer be reduced. To overcome this limitation, it is often necessary to combine them with less local contractors, such as 3B contractors, interval Newton-based contractors, or linear-relaxation-based contractors. Accordingly, the latter three types of contractors are more computationally expensive.
    Interval Newton-based contractors and linear-relaxation-based contractors are efficient only when the box is sufficiently small or when the system is approximately linear, as they typically involve the use of an interval Jacobian matrix or similar structures. If the entries of this interval matrix have large widths, the power of the contractors will be substantially reduced.
    
    Given the varying applicability and computational overhead of these contractors, an effective combination strategy of the contractors is essential for achieving high algorithmic efficiency.
    A straightforward and commonly adopted heuristic is to first apply cheaper contractors and then invoke the more expensive ones~\citep{vu2009enhancing,beelitz2005symbolic,granvilliers2001progress,benhamou1999revising,granvilliers2001combination}.
    In particular, \citet{benhamou1999revising} proposed a classical algorithm that performs constraint propagation, which first employs forward-backward propagation and then applies box consistency techniques only on variables that occur multiple times within the constraint.
    Existing solvers generally adopt similar heuristic for combining the contractors.
%    {\tt{Numerica}}~\citep{van1997numerica} iteratively performs constraint propagation using box consistency, followed with interval-Newton operators, until a sufficient precision is reached.
    The default strategy in {\tt RealPaver}\citep{granvilliers2006algorithm} integrates the method proposed by\citet{benhamou1999revising} with the Hansen–Sengupta operator to enhance pruning efficiency.
    In {\tt IbexSolve }~\citep{chabert2009contractor}, the default strategy employs a sequence of contractors, including the forward-backward propagator, the ACID contractor, an interval Newton-type operator, and a linear-relaxation-based contractor, typically applied in the order presented above.
    Unlike the fixed contractor combinations used in previous works, \citet{araya2015adaptive} proposed an adaptive strategy that fine-tunes the invocation frequency of each contractor. The key idea is to reduce the usage of expensive contractors unless they prove effective in reducing a recently processed box.
    
    In addition to the choice and scheduling of various contractors,
    the implementation parameters or details within individual contractors can also have a considerable effect on algorithmic efficiency, particularly for constraint propagation contractors.
    Constraint propagation contractors, based on hull or box consistency, handle one constraint at a time. 
    When applied to a set of constraints, they must be propagated across all constraints repeatedly until a fixed box is reached, that is, until no further reduction is possible.
    In practice, to balance computational cost and pruning strength, the propagation process may be terminated early when no substantial improvement is observed.
    The appropriate threshold for termination is typically  chosen based on empirical performance~\citep{baharev2011computing}.
    Box-consistency contractors offer more flexibility because they operate on constraint-variable pairs, and each constraint may involve multiple variables. However, applying the contractor to all variables can be costly.
    To mitigate this cost, \citet{goualard2008reinforcement} proposed a dynamic selection strategy that assigns and continuously updates weights for constraint-variable pairs based on their past domain reduction performance, prioritizing those with higher weights.
    Focusing on a single box-consistency contractor, \citet{goldsztejn2010box} proposed an adaptive shaving process that tightens the domain inward from both bounds, yielding more robust performance than the original implementation.
    In~\citet{chabert2007cid}, experiments were conducted to evaluate how different parameter settings in the CID and 3BCID contractors affect overall efficiency on benchmark problems, and to choose appropriate parameters.

    \subsubsection{Strategy on the bisectors}
    So far, we have not discussed bisectors separately, as the operation itself is relatively straightforward—dividing a box into two sub-boxes along a chosen dimension. However, the choice of which dimension to split can still have a significant impact on the efficiency of the algorithm, and determining the optimal choice is a challenging and open question~\citep{baharev2011computing,granvilliers2012adaptive}.
    
    Round-Robin (RR), MaxDom (or Largest-First ) and Smear-based heuristic are among the most commonly adopted heuristic strategies.
    The Round-Robin (RR) strategy selects variables in a circular order; it is simple, deterministic, and does not rely on any problem-specific information or heuristics. 
    The MaxDom strategy selects the variable with the largest domain in the current box, aiming to reduce variable dependency as much as possible, which can lead to stronger contractions through local consistency techniques.
    Smear-based strategies use information of the system and the current box to detect the variable that has the most impact on the system.
    Given a box $\bfx = (\bfx_1, \cdots,  \bfx_n)$, the smear value of $x_i$ with respect to $f_j$ is given by
    $$\smear(x_i,f_j) = |\bfA_{ij}|\cdot wid(\bfx_i)$$
    where $\bfA_{ij}$ is the evaluation of an interval form of 
    $\frac{\partial f_j}{\partial x_i}$ over $\bfx$.
    The MaxSmear heuristic selects the variable $x_i$ which maximizes
    $\max_{j=1,\ldots,n}(\smear(x_i,f_j))$;
    the SumSmear heuristic selects the variable $x_i$ which 
    maximizes
        $\sum_{j=1}^n(\smear(x_i,f_j))$.
    As an improvement, \citet{trombettoni2011inner} proposed heuristics based on the relative smear criterion, which normalize $\smear(x_i,f_j)$ by the sum of all the smear values related to $f_j$.
    The improved heuristics are show to be more efficient than the original ones.
    More sophisticated  smear-based heuristics are presented in~\citet{araya2013more}.
    
    As noted in~\citet{granvilliers2012adaptive}, the efficiency of a subdivision algorithm depends on the coupling of bisection strategy and the contraction procedures.
    Based on the observation that MaxSmear is more useful when the size of the box is small, \citet{granvilliers2012adaptive} proposed an adaptive strategy that favors RR in the early stages of the search,gradually shifts toward MaxSmear, aiming to balance the benefits of both approaches.
    
    An alternative strategy used in the literature is to select the splitting variable based on the output of the contractor.
    In the application of the Hansen-Sengupta operator, the diagonal entries of the interval Jacobian matrix may contain zero. Division by such an interval results in two disjoint intervals separated by a gap. This phenomenon may indicate the presence of multiple solutions. Therefore, when such a situation arises, a reasonable splitting strategy is to exploit the gap for further branching~\citep{hansen1983interval}. 
    Moreover, \citet{batnini2005mind} showed that gaps can occur in hull consistency and box consistency techniques due to division by intervals containing zero. Therefore, splitting strategies that exploit such gaps can  be designed accordingly.
    An analogous splitting strategy is also proposed in~\citet{chabert2007cid}, based on the gaps produced by the CID contractor.
    
    Beyond the classical strategies and their variants, several exploratory efforts have been made to devise novel bisection strategies.
    In addition to local information, as used in MaxDom and smear-based heuristics, \citet{reyes2014probing} attempts to guide the variable selection using past information. This information is gathered in a preprocessing phase through exploratory runs in different directions. After this phase, each variable is assigned a weight, which is then incorporated into the variable selection heuristic during the search process.
    \citet{huang2019evaluating} introduced the lookahead branching heuristic in the attempt to make locally optimal choice.
    It performs exhaustive splitting over each of the variables to estimate the gain from bisection in different directions, and keeps the branch which yields the most progress in the next several steps. 
    The strategy is compared with MaxDom and a smear-based heuristic in the implementation. Interestingly, the experimental results show that different branching methods perform quite differently on larger instances; however, no single heuristic consistently outperforms the others.
    
     \subsection{Other techniques}
    Aside from various contractors, which are the main ingredients of the subdivision algorithm, there are additional techniques that can enhance its efficiency. Here we briefly discuss about the  directed acyclic graph (DAG) representation of constraint system, and the symbolic techniques.
    
    As noted earlier, forward–backward propagation is a basic contractor in subdivision algorithms. However, propagating each constraint \emph{individually} can limit pruning efficiency. To address this, \citet{neumaier2005dag} proposed representing the constraint system as a DAG rather than as separate trees. Common subexpressions across different constraints then appear only once in the DAG, which reduces dependency effects and thereby decreasing interval overestimation.
    This representation yields tighter contractions in forward–backward propagation.
    \citet{vu2008dag,vu2009enhancing} further developed techniques based on this representation and proved its efficiency. 

    Another technique to enhance efficiency is symbolic computation. As noted by \citet{beelitz2005symbolic,benhamou2006continuous}, a nonlinear system can be reformulated by introducing a shared auxiliary variable for each subterm common across different constraints, an idea akin to the DAG representation, in order to mitigate dependency and reduce interval overestimation. 
    More aggressively, additional variables can be introduced to reduce the number of occurrences of each variable within a constraint \citet{granvilliers2001symbolic}, which typically yields tighter contractions at the cost of higher computational effort.
    Symbolic computation can also normalize a polynomial system to an equivalent system with the same zero set, via Gröbner bases or other elimination methods, in the spirit of Gaussian elimination for linear systems. The resulting system is typically easier to compute with. However, Gröbner-basis computation is expensive (with worst-case doubly–exponential complexity), so determining how aggressively to apply symbolic preprocessing is itself a nontrivial design choice~\citep{benhamou2006continuous}.

\section{Dataset description}
Our dataset consists of square systems of nonlinear equations with zero-dimensional solution sets, which are particularly well suited to subdivision-based solvers. It can be divided into two parts: the first part comprises benchmarks drawn from the literature and publicly available datasets, while the second part consists of examples constructed from selected parametric systems arising in practical applications.
The first part of the dataset comprises 581 systems, including 451 polynomial and 130 non-polynomial systems. 
{The second part consists of examples generated from 5 families of parametric systems; by instantiating parameters in each family, we obtain 2,000–30,000 zero-dimensional systems per family.
In what follows, we will describe the structure of the dataset, then present statistics for each part, including system size, number of solutions, and other salient features.}
%    Provide the source of the examples in the dataset and statistical information about the dataset, including the number of variables, the number of solutions, and how many solutions are exact.
%    Describe the forms of the dataset.

\subsection{Overview of the dataset}

The {\bf S}ub{\bf D}ivision {\bf D}ataset (SDD) is organized in the following manner. Inside the root directory SDD, there are two directories: {\em non-parametric} and {\em parametric}. The  {\em non-parametric} directory contains the directory {\em non-polynomial} and the directory {\em polynomial}. Each collected non-parametric system is assigned to a subdirectory with a unique identifier and placed into the above two directories according to its type. Inside the subdirectory for each system lie in four files: the file sys.txt stores the information of the input system; the file output.txt stores the computation time, the number of certified and uncertified solutions, and the number of cells (applicable to subdivision solvers) for three solvers; the file solution.txt stores the solutions of the input system and the solvers used for obtaining such solutions; the file info.txt contains the origin information of the system. 

Inside the {\em parametric} directory, 
each parametric system is assigned to a unique subdirectory, inside which there are three files: 
parameter.txt listing different values of parameters, parametricSys.txt describing the parameter system, and a folder {\em instances} storing the specialized systems with values in parameter.txt. 
Each specialized system is now a non-parametric one and is organized exactly the same as the ones 
in the  {\em non-parametric} directory.

The solutions of the systems are obtained with two subdivision solvers and a symbolic solver on a desktop. 
The precise information of the software and hardware is reported in Table~\ref{tab:info}. 
%{\color{blue}
For all the solvers, we impose a timelimit of 1000 seconds.
 IbexSolve is called with the option ``ibexsolve -e 1e-6", which sets the bisection tolerance to $10^{-6}$. 
 RealPaver is called with the option ``-number +oo -precision 1e-6", the reason for using this option will be detailed in~\ref{subsec:application1}.
 %}
 
\begin{table}[htbp]
    \centering
    \caption{Hardware and software information.}
    \label{tab:info}
    \begin{tabular}{ll}
    \toprule
     Machine    &  Intel i9-11900K@3.50GHz, MEM 128GB, Ubuntu 20.04\\
     IbexSolve     & Release 2.8.9, with option ``-e 1e-6" \\
     RealPaver    &  v. 0.4 (c) LINA 2004, with option ``-number +oo -precision 1e-6" \\
     RootFinding & Maple 2025 \\
     \bottomrule
    \end{tabular}
\end{table}

\subsection{Collection of existing examples from the literature}
    We report the details and summary statistics for the first part of the dataset, which consists of benchmarks collected from existing literature. The examples were gathered from a variety of sources, including  the IbexSolve~\citep{chabert2009contractor} and RealPaver~\citep{granvilliers2006algorithm} software suites, the COCONUT~\citep{shcherbina2002benchmarking}, PHCpack~\citep{verschelde2011polynomial}, and ALIAS benchmarks~\citep{merlet2007alias}, as well as numerous published references, such as~\citet{chen14real,lukvsan1999sparse,stuber2010nonsmooth,van1997solving,wang1996,vu2005rigorous,xia2006real,abbott1975fast,HOU2019145,mantzaflaris2011continued,sherbrooke1993computation,verschelde1994homotopy,KuriMorales2002SolutionOS,Moore2009Introduction,gawali2022integrative}, etc.
    When collecting the examples, an initial screening has been made to discard obviously duplicate examples.
    This results in a total of 770 examples  listed in Table~\ref{tab:source}.
After solving these examples with subdivision methods, by examining the examples with the same number of solutions
and carefully comparing the input systems, we further identify nearly 200 duplicate examples, 
including those with variables renamed. Finally we obtain a dataset containing 451 polynomial systems and 130 non-polynomial systems.
    We believe  that these examples cover the vast majority of cases reported in the literature. Their sources span a wide range of application domains, including economics, chemical reaction kinetics, circuit analysis, mechanical design, computer vision, neurophysiology, and more.

\begin{table}[htbp]
    \centering
        \caption{Sources of examples.}
    \label{tab:source}
    \small
    \renewcommand{\arraystretch}{0.85}
    \setlength{\tabcolsep}{6pt}
    \begin{tabular}{l r l r}
    \toprule
    \multicolumn{2}{c}{\textbf{Source}} & \multicolumn{2}{c}{\textbf{Source}} \\
    \cmidrule(r){1-2} \cmidrule(l){3-4}
    \textbf{Name} & \textbf{\#Ex.} & \textbf{Name} & \textbf{\#Ex.} \\
    \midrule
    IbexSolve benchmark\tablefootnote{\url{https://ibex-team.github.io/ibex-lib/}} & 260 &
    PHCpack database\tablefootnote{\url{https://homepages.math.uic.edu/~jan/demo.html}} & 85 \\
    COCONUT benchmark\tablefootnote{\url{https://arnold-neumaier.at/glopt/coconut/Benchmark/Library3_new_v1.html}} & 101 &
    ALIAS benchmark\tablefootnote{\url{https://www-sop.inria.fr/coprin/logiciels/ALIAS/Benches/benches.html}} & 104 \\
    galaad\tablefootnote{\url{https://www-sop.inria.fr/galaad/data/BASE/2.multipol/centralpos.html}} & 9 &
    \citet{chen14real} & 51 \\
    \citet{lukvsan1999sparse} & 22 &
    RealPaver benchmark\tablefootnote{\url{https://realpavergroup.univ-nantes.io/realpaver/}} & 18 \\
    \citet{stuber2010nonsmooth} & 17 &
    \citet{van1997solving} & 21 \\
    \citet{wang1996} & 16 &
    \citet{vu2005rigorous} & 16 \\
    \citet{xia2006real} & 6 &
    others & 44 \\
    \midrule
    \multicolumn{3}{r}{Total} & 770 ({\bf 581}) \\
    \bottomrule
    \end{tabular}

\end{table}

Table~\ref{tab:poly_statistics} summarizes the distribution of the polynomial systems in our benchmark with respect to the number of variables. For each variable range, it reports the number of examples, together with the ranges of numbers of equations and terms, the total and partial degrees, the coefficient magnitudes, as well as the number of instances solved by each solver.

\begin{table}[htbp]
\centering
\caption{Statistics of the polynomial problems.}
\label{tab:poly_statistics}
\renewcommand{\arraystretch}{1.15}
\setlength{\tabcolsep}{4pt}
\scriptsize
\begin{tabular}{cccccccccc}
\toprule
\multicolumn{7}{c}{Problem characteristics} & \multicolumn{3}{c}{Solved instances} \\
\cmidrule(lr){1-7} \cmidrule(lr){8-10}
\makecell[l]{Variables} & 
    \makecell{Examples} & 
    \makecell{Equations} & 
    \makecell{Terms} & 
    \makecell{Total \\ Degree} & 
    \makecell{Partial \\ Degree} &   
    \makecell{Coefficients \\ Range} &
    \makecell{IbexSolve} &         
    \makecell{RealPaver} &
    \makecell{Maple} \\
    \midrule
        1--3       & 93  & 1-3      & 1-61   & 1-21 & 1-21  & $[-1.38\times 10^{19},\,\ 1.38\times 10^{19}]$ & 92 & 108 & 93\\
        4--6       & 104 & 4-6      & 2-32   & 2-11 & 1-6   & $[-1.92\times 10^{7},\,\ 1.59\times 10^{14}]$ & 86 & 75 & 93\\
        7--10      & 100 & 7-10     & 2-64   & 1-13 & 1-7   & $[-1.21\times 10^{15},\,\ 8.26\times 10^{14}]$ & 87 & 64 & 70\\
        11--25     & 76  & 11-25    & 2-96   & 2-30 & 1-30   & $[-200,\, 220]$ & 55 & 32 & 18\\
        26--50     & 41  & 26-44    & 1-121  & 2-30 & 1-3    & $[-400,\, 293.0]$ & 24 & 10 & 4\\
        51--100    & 13  & 51-100   & 1-83   & 2-3  & 2-3    & $[-80.0,\, 12.8]$ & 9  & 6  & 0\\
        101--500   & 13  & 101-482  & 1-403  & 3-4  & 2-3    & $[-400.0,\, 12.8]$ & 6  & 4  & 0\\
        501--1000  & 10  & 501-1000 & 3-901  & 2-3  & 2-3    & $[-900.0,\, 11.8]$ & 0  & 1  & 0\\
        $>1000$    & 1   & 1001& 5-1001 & 3  &3    & $[-1000.0,\, 11.8]$ & 0  & 0  & 0\\
    \bottomrule
\end{tabular}
\end{table}

Table~\ref{tab:stat_nonpoly} provides an overview for the 130 non-polynomial systems. It classifies the systems according to the types of non-polynomial functions involved and for each class reports the number of examples together with the ranges of numbers of variables and terms.
\begin{table}[htb]
    \centering
    \caption{Information on non-polynomial systems.}
    \label{tab:stat_nonpoly}
    \small
    \setlength{\tabcolsep}{3pt}      % 缩小列间距
    \renewcommand{\arraystretch}{0.9}% 略微压缩行高
    \begin{tabular}{@{} l r c c @{}}
        \toprule
        Type & \# Ex. & No. of variables & No. of terms \\
        \midrule
        Rational functions % $x/y$
        & 5  & 1-9    & 2-7    \\
        Radicals (algebraic functions) % $x^{1/2}$
        & 3 & 1    & 3-7    \\
        Exponential functions % $e^x$
        & 38 & 1-1000 & 2-6    \\
        Logarithmic functions % $\ln(x+y)$ 
        & 1 & 1    & 2    \\
        Trigonometric functions % $\sin(x)$ 
        & 49 & 1-1001 & 2-1001 \\
%        Inverse trigonometric functions %$\arcsin(x)$ 
%        & 0 & 0 & 0 \\
        Absolute value function % $\|x|$
        & 1 & 2  & 2   \\
        Hyperbolic function % $\cosh(x)$
        & 3 & 1-2  & 3-4   \\
        Mixed forms of the above % $\sin(x)e^x$
        & 30 & 1-100  & 2-12   \\
        \bottomrule
    \end{tabular}
    
\end{table}

% \begin{figure}[h!]
%     \centering
%     \includegraphics[width=0.5\textwidth]{task5/poly_term_log.png} 
%     \includegraphics[width=0.5\textwidth]{task5/poly_degree.png} 
%     \caption{Statistics of the polynomial benchmarks}
%     \label{fig:poly_statistic}
% \end{figure}

Table~\ref{tab:num_solutions} reports the distribution of the number of certified solutions for polynomial and non-polynomial systems. A system is included in this table only if at least one of IbexSolve or Maple returns an exact root count, so that no suspect regions remain. Among the 451 polynomial and 130 non-polynomial systems in our benchmark, the solvers succeed in determining the exact number of solutions for 358 and 91 systems, respectively. 
%{\color{red}
For the remaining instances, none of the solvers manages to certify the root count within 1000 seconds, either because they all hit the time limit, crash or error out, or because their outputs still contain suspect regions that may hide a root.
%}

\begin{table}[htbp]
  \centering
   \caption{Distribution of the number of solutions.}
  \label{tab:num_solutions}
  \small
  \begin{tabular}{lrr|lrr}
    \toprule
    \# Zeros & \# Poly & \# Non-poly & \# Zeros & \# Poly & \# Non-poly\\
    \midrule
    0      & 8 & 11 &  11--20      & 35 &  2 \\
    1      & 80 & 33 &  21--50      & 28 &  5 \\
    2      & 71 & 24 &  51--100     &  4 &  2 \\
    3      & 25 &  5 &  101--200    &  8 &  1 \\
    4      & 26 &  3 &  201--1000   &  3 &  0 \\
    5      &  8 &  0 &  $>1000$     &  1 &  0 \\
    6--10  & 61 &  5 \\
      \bottomrule
  \end{tabular}
 
\end{table}

% \begin{table}[htbp]
%   \centering
%   \caption{Distribution of the number of solutions.}
%   \label{tab:num_solutions}
%   \small
%   \begin{tabular}{lrr}
%     \toprule
%     Number of solutions & \# Polynomial & \# Non-polynomial \\
%     \midrule
%     0           & 12 & 16 \\
%     1           & 85 & 29 \\
%     2           & 69 & 20 \\
%     3           & 25 &  4 \\
%     4           & 26 &  2 \\
%     5           &  8 &  0 \\
%     6--10       & 54 &  5 \\
%     11--20      & 36 &  2 \\
%     21--50      & 24 &  5 \\
%     51--100     &  4 &  2 \\
%     101--200    &  8 &  1 \\
%     201--1000   &  4 &  0 \\
%     $>1000$     &  1 &  0 \\
%     \bottomrule
%   \end{tabular}
 
% \end{table}

%     These examples typically originate from over ten references such as $VanHentenryck-1997-SPS$, $moore2009introduction1$, $improvedKrawcyzk2019$, etc., as well as benchmarks from software like IBEX and RealPlayer, and databases made public by the COPRIN project group at the French National Institute for Research in Computer Science and Automation (INRIA), which cover a range of nonlinear equation systems from various fields and complexities, ensuring their representativeness and diversity.

\subsection{Parametric Systems}  
In this section, we present five class of parametric systems in the dataset.
Table~\ref{tab:parametric} summarizes their key information, including the system type and the numbers of variables and parameters.
For each class, we instantiate the parameters to generate a collection of zero-dimensional square systems.
Table~\ref{tab:parametric} reports, for each class, the number of instances that IbexSolve fails to solve within the 1000-second time limit. For the remaining instances, Table~\ref{tab:parametric} further reports the average solving time.
  
\begin{table}[ht]
  \centering
  \scriptsize
  \caption{Summary of parametric systems.}
  \label{tab:parametric}
  \setlength{\tabcolsep}{3pt} % 列间距稍微缩小一点
  \begin{tabular}{|c|c|c|c|c|c|c|c|}
    \hline
    \makecell[c]{Application} &
    \makecell[c]{System} &
    \makecell[c]{Type} & 
    \makecell[c]{Number of\\Variables\\(Equations)} &
    \makecell[c]{Number of\\Parameters} &
    \makecell[c]{Number of\\Instances} &
    \makecell[c]{Average\\Timing (s)} &
    \makecell[c]{Number of\\Timeout\\Instances} \\
    \hline

    \multirow{4}{*}{\makecell[c]{Multi-joint\\Robot Arm}} &
    \makecell[c]{Planar-\\Trigonometric} &
    Non-polynomial & 30--42 & 8--10 & 5000 & 9.33 & 65 \\
    \cline{2-8}
    & \makecell[c]{Planar-\\Polynomial} &
    Polynomial & 20--28 & 8--10 & 5000 & 2.84 & 99 \\
    \cline{2-8}
    & \makecell[c]{Spatial-\\Trigonometric} &
    Non-polynomial & 9--15 & 3--7 & 3000 & 9.55 & 303 \\
    \cline{2-8}
    & \makecell[c]{Spatial-\\Polynomial} &
    Non-polynomial & 15--25 & 3--7 & 3000 & 15.6 & 389 \\
    \hline

    % ---- Stewart platform ----
    \makecell[c]{Stewart\\Platform} &
    Stewart &
    Polynomial & 9 & 29 & 10000 & 1.04 & 0 \\
    \hline

    % ---- Kuramoto model ----
    \makecell[c]{Kuramoto\\Model} &
    Kuramoto &
    Polynomial & 10 & 5 & 10000 & 0.351 & 0 \\
    \hline

    % ---- Flah unit ----
    \makecell[c]{Flash Unit} &
    Flash unit &
    Non-polynomial & 28 & 4 & 10000 & 7.80 & 0 \\
    \hline

    % ---- Orbit determination ----
    \makecell[c]{Orbit\\Determination} &
    Orbit &
    Polynomial & 14 & 30 & 2000 & 358 & 277 \\
    \hline
  \end{tabular}
\end{table}

%\subsubsection{Randomly Generated Robotic Examples}
\subsubsection{Multi-joint Robot Arm}\label{subsec:robot}
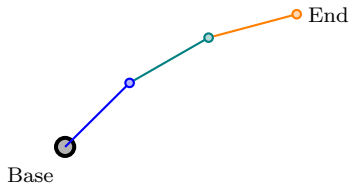
\begin{figure}[h]
    \centering
    \begin{tikzpicture}[thick, scale=0.8]
        % Base
        \filldraw[fill=gray!60, draw=black, line width=1.5pt] (0,0) circle (0.15);
        \node[font=\scriptsize, anchor=north east] at (0,-0.15) {Base};

        % Arms (longer)
        \draw[thick, draw=blue] (0,0) -- ++(45:1.5) coordinate (A);
        \draw[thick, draw=teal] (A) -- ++(30:1.5) coordinate (B);
        \draw[thick, draw=orange] (B) -- ++(15:1.5) coordinate (C);

        % Joints
        \filldraw[fill=blue!30, draw=blue] (A) circle (0.07);
        \filldraw[fill=teal!30, draw=teal] (B) circle (0.07);
        \filldraw[fill=orange!30, draw=orange] (C) circle (0.07);

        % End node label
        \node[font=\scriptsize, anchor=west] at (C) {End};
    \end{tikzpicture}
    \caption{Multi-joint robot arm.}
    \label{fig:arm}
\end{figure}
    Multi-joint robot arms are widely used in robotics for tasks such as manipulation, assembly, and precise positioning. A fundamental problem arising in their analysis is the localization of planar or spatial multi-joint robot arms: given the coordinates of the end point (end effector), one seeks to determine the feasible configurations of the intermediate joints. This localization naturally leads to a system of nonlinear equations derived from the kinematic constraints of the robot arm.
    
    We focus on the localization problem of a simple serial multi-joint robot arm.
    Assume the robot consists of $m$ serially connected links $L_1, \dots, L_m$ extending from the base to the end effector, with respective lengths $l_1, \dots, l_m$. 
    The base is fixed at the origin. 
    
First, we consider the planar case of the localization problem. 
Given the coordinates of the end point of the last link $(x_m, y_m)$, 
the goal is to determine the feasible positions and orientations of all intermediate joints. 
In this setup, the resulting system of kinematic equations has a total of $3m-2$ unknowns: 
$(x_i, y_i)$ for $i = 1, \dots, m-1$, representing the end point coordinates of the $i$-th link, 
and $\theta_i$ for $i = 1, \dots, m$,  representing the orientation of the $i$-th link measured counterclockwise from the positive $x$-axis.
This system consists of $2m$ equations:
$$
P_{1}=\left\{
\begin{array}{rcl}
x_j - \sum_{i=1}^j \ell_i \cos(\theta_i) & = & 0, \\
y_j - \sum_{i=1}^j \ell_i \sin(\theta_i) & = & 0,
\end{array}
\right.
\qquad j=1,\ldots,m.
$$
    With the coordinates $(x_m,y_m)$ fixed, this system contains $3m-2$ variables and $2m$ equations, so for $m \ge 3$ the solution set is generally positive-dimensional. 
    
To obtain a zero-dimensional system, we need to fix $m-2$ variables.
%, selected from $(x_i, y_i)$ for $i=1,\dots,m-1$. 
Randomly assigning values to these variables would most likely lead to an inconsistent system with no solution. 
Therefore, the system needs to be constructed with care.
% {\color{red} Remove this part?
% To avoid this, we first construct a consistent system in which all $(x_i, y_i)$ values are known. 
% We then fix $m-2$ of the immediate coordinates as long as the coordinates of the final end as constants, while keeping the remaining ones and all $\theta_i$ for $i=1,\ldots,m$ as unknowns, resulting in a system of $2m$ equations with $2m$ unknowns.
% % During this process, care must be taken in selecting which variables to fix, ensuring that any subsystem formed by a consecutive subset of the links is neither overdetermined nor underdetermined.
% }
%
We now describe in detail how the systems are generated. 
First, the number of links $m$ is fixed, and the lengths of the links are randomly chosen as integers from 1 to 10. 
Next, we construct a consistent system by randomly assigning $\theta_i \in [-\pi, \pi]$ for $i=1,\dots,m$ and computing the values of the coordinates $(x_i, y_i)$ for $i=1,\dots,m$. 
Now we have a known consistent system, we construct a zero-dimensional square system by fixing the end point coordinates 
$(x_m, y_m)$ of the last link and
$m-2$ coordinates in $\{x_i, y_i\}_{i=1,\ldots, m-1}$ as constants, and treating the remaining variables as unknowns.
The unknowns consist of $\{\theta_i\}_{i=1,\ldots,m}$ and $m$ variables in $\{x_i, y_i\}_{i=1,\ldots, m-1}$. There are $2m$ unknowns and $2m$ equations.
The intervals of interest for the variables are set as $x_j, y_j \in [-\sum_{i=1}^j \ell_i, \sum_{i=1}^j \ell_i]$ and $\theta_j \in [-\pi, \pi]$ for $j=1,\dots,m$.

 During the process of fixing $m-2$ coordinates in $\{x_i, y_i\}_{i=1,\ldots,m-1}$, care must be taken in choosing which variables to fix so that the resulting system has a zero-dimensional but non-empty solution set. The basic idea is to ensure that any subsystem formed by a consecutive subset of links is neither overdetermined nor underdetermined. The implementation is as follows:
 1) Randomly select $m-2$ coordinates from the set $\{x_i, y_i\}_{i=1,\ldots, m-1}$.
 2) Arrange all selected coordinates into an ordered list $\mathcal{L}$ by sorting them in the order 
 $x_1< y_1< x_2<y_2 <\cdots < x_{m-1} < y_{m-1}$.
 %first by the link index $i$, and, for each fixed index, placing $x_i$ before $y_i$ in lexicographic order. 
 3) For each index $j$ such that both $x_j$ and $y_j$ appears in $\mathcal{L}$, check whether the number of elements preceding $x_j$ in $\mathcal{L}$ is equal to $j-2$. If this condition holds, then the subsystem formed by the links from $1$ to $j$ is a square system; otherwise, the current random selection is rejected.
We repeat this random selection–and–checking procedure until the chosen coordinates satisfy the above conditions.
We use Python scripts to generate 5000 instances in total, with $m$ ranging from 10 to 14, and 1000 instances generated for each value of $m$. 
The range of $m$ is chosen to ensure that the solving time for the systems is neither excessively long nor trivially short.

%     Use Python scripts to randomly generate datasets for solving nonlinear systems involved in motion planning problems for multi-body robots. During the generation process, different numbers of variables and equation forms are considered to ensure the diversity and representativeness of the dataset. Each generated equation system records the number of solutions and runtime for further analysis. The specific steps for generating are as follows:
% \begin{itemize}
%     \item Define the variable range: 
%     \begin{itemize}
%         \item The value of $m$ is chosen from [1, 5];
%         \item each $li$ takes a random integer value within the range of 1 to 10;
%         \item each takes a random integer value within its range. 
%     \end{itemize}
%     \item Generate equations: Specifically includes two types of planar rotating multi-body robots and one type of spatial multi-body robot motion planning problems.
%     \item Validate the equation system: Check if the generated equation system meets the screening criteria and if it has solutions.
% \end{itemize}

Beyond expressing the problem as the nonlinear system described above, we also formulate the same problem in polynomial systems, for comparison.
It suffices to replace $ \sin(\theta_i) $ and $ \cos(\theta_i) $ in $P_{1}$ with auxiliary variables $s_i$ and $c_i$, respectively, and to add the constraints $c_i^2 + s_i^2 = 1$ for $i = 1, \ldots, m$.
%{\color{blue} This yields a polynomial system with $3m$ equations and $4m-2$ variables, of which $m-2$ need to be fixed, as before.}
The procedure for generating random instances is the same as described above: first construct a consistent system, and then fix $(x_m,y_m)$ and $m-2$ of the coordinates in $\{x_i, y_i\}_{i=1,\ldots,m-1}$ as constants. The only difference is that the unknown $\{\theta_i\}_{i=1,\ldots,m}$ are replaced by $\{s_i,c_i\}_{i=1,\ldots,m}$, together with the additional constraints $\{c_i^2 + s_i^2 = 1\}_{i = 1, \ldots, m}$. 
This yields a polynomial system with $3m$ equations and $3m$ variables.
We generated 5000 instances, representing the same problems as described above, except that they are expressed as polynomial systems.
When solving the systems, the intervals of interest for $c_i$ and $s_i$ are set to $[-1,1]$.

Each trigonometric instance in our benchmark has a matching polynomial counterpart: the two formulations encode the same underlying problem, differing only in representation. We solve both versions with IbexSolve using ``ibexsolve -e1e-6", which sets the bisection tolerance to $10^{-6}$. With this configuration, we run IbexSolve on the two formulations independently and record the outcomes.

Across the 5,000 instances, IbexSolve reports 196 distinct root counts, ranging from 2 to 1,920.
Table~\ref{tab:rootcount-robot1} summarizes the distribution of root counts for different numbers $m$ of links. We aggregate results from both the trigonometric and polynomial formulations: an instance is included as long as IbexSolve returns an exact root count for at least one formulation. Instances are excluded if, for both formulations, IbexSolve either fails to terminate within the 1,000 seconds time limit or returns suspect regions.
For this class of instances, all excluded cases are due to timeouts in both formulations.
In the table, the column ``Exact" reports the number of instances for which an exact root count is available, the column ``Min" (resp. ``Max") reports the minimum (resp. maximum) number of roots, and the column ``C" reports the number of distinct root counts observed. The remaining columns report, for each root-count range, how many instances fall into that interval.

% TODO: An analysis for the failed examples (timeout, not reliable (Jacobian?), bug (re-randomize and replace, Lp-relaxation))? 
% What about the running time?
% Are the number of roots also power of 2?
% Add columns on timeout and inexact cases and merge 
% the root count intervals?
\begin{table}[hbt]
\centering
\small
\setlength{\tabcolsep}{5pt}
\caption{Distribution of exact root counts in the planar case.}
\label{tab:rootcount-robot1}
\begin{tabular}{r|r|rrr|rrrrrrr}
\hline
$m$ & Exact & Min & Max & C & 1--10 & 11--50 & 51--100 & 101--200 & 201--500 & 501--1000 & $>1000$ \\
\hline
10 & 999 & 2 & 272  & 73  & 150 & 583 & 186 & 69  & 11  & 0  & 0 \\
11 & 998 & 2 & 556  & 87  & 102 & 498 & 247 & 115 & 34  & 2  & 0 \\
12 & 997 & 2 & 896  & 108 & 44  & 440 & 254 & 175 & 71  & 13 & 0 \\
13 & 993 & 4 & 1664 & 107 & 34  & 307 & 271 & 213 & 137 & 28 & 3 \\
14 & 998 & 4 & 1920 & 116 & 21  & 251 & 242 & 233 & 186 & 57 & 8 \\
\hline
\end{tabular}
\end{table}

From Table~\ref{tab:rootcount-robot1}, we observe that IbexSolve 
can reliably isolate the right number of real roots for most of the instances. As $m$ increases, the number of distinct feasible configurations for the multi-joint robot arm substantially increases. 
%} % color read

In addition to the planar multi-joint robot arm, we also consider the localization problem for spatial multi-joint robot arms. The setup is the same as above, except that the links move in three-dimensional space rather than in a plane.
The problem can be formulated as 
$$
    P_1'=\left\{
    \begin{array}{rclr}
       x_j - \sum_{i=1}^j\ell_i cos(\theta_i)cos(\phi_i) & = & 0, & for~ j=1,\ldots,m\\
       y_j - \sum_{i=1}^j\ell_i cos(\theta_i)sin(\phi_i) & = & 0, & for~ j=1,\ldots,m\\
       z_j - \sum_{i=1}^j\ell_i sin(\theta_i) & = & 0, & for~ j=1,\ldots,m\\
    \end{array}
    \right.
    $$
where $(x_i,y_i,z_i)$ for $i=1,\ldots,m$ denotes the coordinates of the end point of link $i$. In practice, $(x_m,y_m,z_m)$ is usually fixed.
Then this system has $5m-3$ variables and $3m$ equations, so for $m\geq2$, $2m-3$ of the variables must be fixed to obtain a zero-dimensional system. 
To avoid generating a system that is likely to be inconsistent, we first construct a known system by randomly assigning values to the variables $\theta_i$ and $\phi_i$ for $i=1,\dots,m$. 
We then fix $2m-3$ coordinates among $(x_i, y_i, z_i)$ for $i=1,\dots,m-1$, and fix the end coordinates $(x_{\rm end}, y_{\rm end}, z_{\rm end})$.
This results in a system of $3m$ variables and $3m$ equations.
We use python script to generate $3000$ instances in total, with $m$ ranging from 3 to 5, and 1000 instances generated for each value of $m$. 
The procedure for generating instances and the criteria for selecting $m$ follow essentially the same line as outlined above.

Just as for the planar systems, we also produce polynomial formulations for the spatial systems.
Specifically, we replace $sin(\theta_i), cos(\theta_i), sin(\phi_1)$ and $cos(\phi_i)$ with four variables $st_i, ct_i, sp_i, cp_i$ respectively, and add the constraints $st_i^2 + ct_i^2 = 1, sp_i^2 + cp_i^2 =1$ for $i=1,\ldots, m$.
We generated $3000$ polynomial systems, each encoding the same underlying problem as its trigonometric counterpart.

As in the planar case, we solve both the trigonometric and polynomial formulations with IbexSolve and record the outcomes. Table~\ref{tab:rootcount-robot3} summarizes the distribution of root counts for these systems.
We aggregate the results across the two formulations in the same way as for the planar systems. 
In the table, the column ``Exact" reports the number of instances for which an exact root count is available. 
For the instances where IbexSolve fails to produce an exact root count in both formulations, almost all cases correspond to timeouts in both formulations. The only exception is a single instance for $m=5$: one formulation times out, while the other terminates but produces non-certified output (i.e., the output includes unknown boxes).
The columns on the right report, for each root count, the number of instances attaining that value.
For this set of instances, the root counts takes fewer distinct values, since the numbers of links $m$ are smaller compared to the planar case.
% {\color{red}
% TODO: the number of roots is always power of 2?
% For m=4 and 5, the number of instances is less than 1000.
% Is this due to timelimit or failure to isolate (how many both timeout and how many at least one reliable)?
% } % color red

\begin{table}[hbt]
\centering
\small
\setlength{\tabcolsep}{5pt}
\caption{Distribution of exact root counts in the spatial case.}
\label{tab:rootcount-robot3}
\begin{tabular}{cccccccccccccc}
\hline
$m$ & Exact & 16 & 32 & 64 & 96 & 128 & 192 & 256 & 320 & 384 & 448 & 512\\
\hline
3 & 1000 & 586 & 414 & 0   & 0  & 0   & 0  & 0   & 0  & 0  & 0 & 0 \\
4 & 943  & 0   & 333 & 426 & 48 & 136 & 0  & 0   & 0  & 0  & 0 & 0 \\
5 & 844  & 0   & 0   & 148 & 0  & 370 & 51 & 205 & 13 & 20 & 7 & 30 \\
\hline
\end{tabular}
\end{table}

%{\color{blue}
We compare the performance of IbexSolve on the trigonometric and polynomial formulations for both the planar and spatial cases. Among the instances where IbexSolve successfully returns a result, we observe that the polynomial formulation has a shorter average solving time in the planar case, whereas the trigonometric formulation is faster on average in the spatial case. However, in both settings, the polynomial formulation exhibits a higher incidence of timeouts and non-certified outcomes (i.e., runs that terminate with unknown boxes).
The details are presented in Table~\ref{tab:ibex-robot}.
%}

% Preamble:
% \usepackage{booktabs}
% \usepackage{makecell}

\begin{table*}[hbt]
\centering
\small
\setlength{\tabcolsep}{3.2pt}
\renewcommand{\arraystretch}{1.15}
\caption{IbexSolve outcomes for robot systems. Each entry reports (top to bottom): \textbf{TO} = \#(timeout), \textbf{NC} = \#(non-certified), and \textbf{Avg} = average time (s) computed over instances not timeout.}
\label{tab:ibex-robot}
\begin{tabular}{l l|ccccc|ccc}
\toprule
& & \multicolumn{5}{c|}{\textbf{Planar case}} & \multicolumn{3}{c}{\textbf{Spatial case}} \\
 & & \textbf{m=10} & \textbf{m=11} & \textbf{m=12} & \textbf{m=13} & \textbf{m=14}
& \textbf{m=3} & \textbf{m=4} & \textbf{m=5} \\
\midrule

\multirow{3}{*}{\textbf{Trig.}}
& \textbf{TO}  & 8   & 12  & 12  & 22  & 11  & 0   & 84  & 219 \\
& \textbf{NC}  & 1   & 0   & 0   & 1   & 0   & 1   & 1   & 0   \\
& \textbf{Avg} & 3.598 & 5.357 & 7.942 & 9.941 & 19.834 & 1.230 & 10.862 & 18.656 \\
\midrule

\multirow{3}{*}{\textbf{Poly.}}
& \textbf{TO}  & 18  & 13  & 18  & 24  & 26  & 23  & 129 & 237 \\
& \textbf{NC}  & 1   & 2   & 0   & 0   & 1   & 2   & 3   & 4   \\
& \textbf{Avg} & 0.630 & 2.024 & 2.657 & 3.903 & 5.013 & 6.673 & 19.626 & 22.517 \\
\bottomrule
\end{tabular}
\end{table*}

%{\color{blue}
Since the trigonometric and polynomial instances encode the same underlying problem, we cross-check the solver outputs by comparing the numbers of certified solution boxes produced for the two formulations. For all instances reported in the tables above, the results of the two formulations are consistent.

However, we found a few planar instances (not included in the table) for which the two formulations yield inconsistent results. For example, for one instance the trigonometric formulation produces 64 certified solution boxes and reports no suspect regions, whereas the corresponding polynomial formulation produces neither certified solutions nor suspect boxes. Moreover, the polynomial run terminates after pruning the initial domain at the very beginning of the search, without any bisection.
Interestingly, if we manually bisect the initial domain into two sub-boxes and solve the two subproblems separately, each subproblem returns certified solutions, and the combined solution count becomes consistent with that of the trigonometric formulation.
This behavior suggests an instability in the solver’s pruning on the polynomial formulation: simply splitting the initial box into two sub-boxes can change the outcome of the solver, even though the underlying set of real solutions is unchanged.
However, we found that disabling the linear-relaxation component makes IbexSolve produce stable results, and the certified solution counts on these problematic instances become consistent with those obtained from the trigonometric formulation.
%}

\subsubsection{Stewart platform}
\begin{figure}[h!]
    \centering
    \includegraphics[
    width=0.45\textwidth]{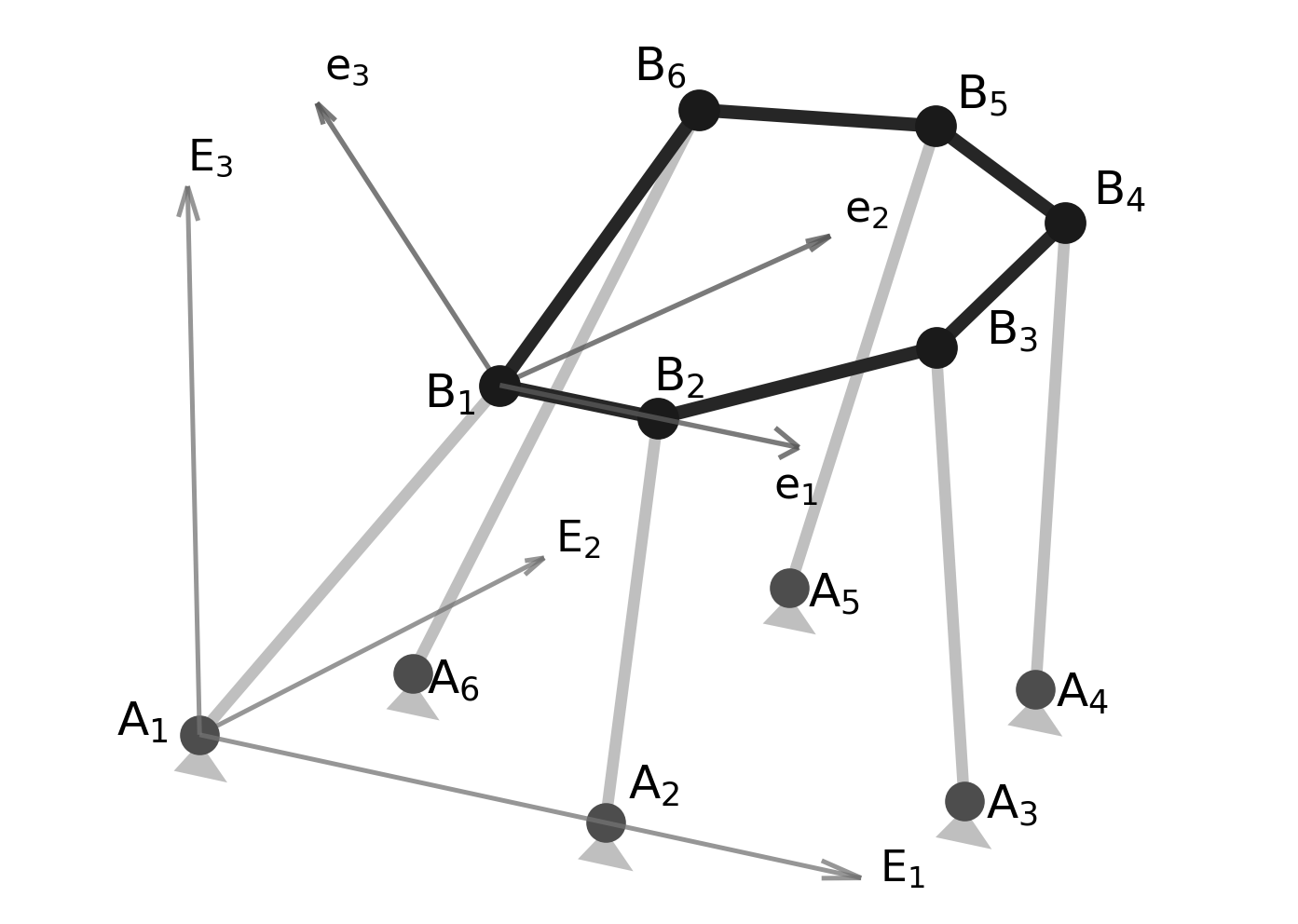} % 调整宽度可选
    \caption{A Stewart platform.}
    \label{fig:stewart}
\end{figure}

The Stewart platform, as show in Figure \ref{fig:stewart} is a parallel robotic mechanism consisting of a movable platform, which is a rigid body, connected to a fixed base by six adjustable-length rods, widely used in flight simulators, precision machining, and motion control.
Given the lengths of the connecting rods, the problem is to determine all possible configurations (postures) of the movable platform. This is a well-known and important problem in robotics, as it is critical for the design, control, and motion planning of parallel manipulators.
\citet{dietmaier1998stewart} showed that this problem can have (up to) 40 solutions, and it has been used as a challenging benchmark in the literature~\citep{baharev2017robust}.

This classic problem can be formulated in various ways. 
Here, we adopt the formulation of \citet{dietmaier1998stewart}, as it provides a relatively clear framework and allows for convenient parameter specification.

Figure~\ref{fig:stewart} illustrates the setup of the Stewart platform. An orthogonal coordinate frame is fixed to the base with origin at $A_1$ and axes $E_1, E_2, E_3$, while another frame is attached to the platform (rigid body) with origin at $B_1$ and axes $e_1, e_2, e_3$. Here $E_1, E_2, E_3$ and $e_1, e_2, e_3$ are vectors. These symbols are typically written in boldface in the literature, but we keep the plain (non-bold) notation here for consistency with the preceding sections.
The directions of $E_1$ and $e_1$ are chosen so that points $A_2$ and $B_2$ lie along these axes, respectively, and $E_2$ and $e_2$ are oriented such that points $A_3$ and $B_3$ lie in the planes defined by $(E_1, E_2)$ and $(e_1, e_2)$, respectively. A configuration is feasible if each corresponding pair of points $A_j$ and $B_j$ is separated by a distance $L_j$, the length of the connecting rod.
 For simplicity, we assume $L_1 = 1$.
 A unit vector $n$ pointing from $A_1$ to $B_1$ is indicated in the figure.
 Let $a_{ij}$ ($b_{ij}$, respectively) denote the coordinates of $A_i$ ($B_i$, respectively) in the base (platform, respectively) frame.
The constraints imposed by the connecting rods can be formulated as:
% \begin{equation}
% \left\{
% \begin{array}{l}
% \| n + b_{21} e_1 - a_{21} E_1 \|^2 = L_1^2, \\
% \| n + b_{31} e_1 + b_{32} e_2 - a_{31} E_1 - a_{32} E_2 \|^2 = L_2^2, \\
% \| n + b_{41} e_1 + b_{42} e_2 + b_{43} e_3 - a_{41} E_1 - a_{42} E_2 - a_{43} E_3 \|^2 = L_3^2, \\
% \| n + b_{51} e_1 + b_{52} e_2 + b_{53} e_3 - a_{51} E_1 - a_{52} E_2 - a_{53} E_3 \|^2 = L_4^2, \\
% \| n + b_{61} e_1 + b_{62} e_2 + b_{63} e_3 - a_{61} E_1 - a_{62} E_2 - a_{63} E_3 \|^2 = L_5^2,
% \end{array}
% \right.
% \label{eq:stewart-rods}
% \end{equation}

\begin{equation}
\left\{
\begin{array}{l}
\| n + b_{21} e_1 - a_{21} E_1 \|^2 = L_2^2, \\
\| n + (b_{31} e_1 + b_{32} e_2) - (a_{31} E_1 + a_{32} E_2) \|^2 = L_3^2, \\
\| n + (b_{41} e_1 + b_{42} e_2 + b_{43} e_3) - (a_{41} E_1 + a_{42} E_2 + a_{43} E_3) \|^2 = L_4^2, \\
\| n + (b_{51} e_1 + b_{52} e_2 + b_{53} e_3) - (a_{51} E_1 + a_{52} E_2 + a_{53} E_3) \|^2 = L_5^2, \\
\| n + (b_{61} e_1 + b_{62} e_2 + b_{63} e_3) - (a_{61} E_1 + a_{62} E_2 + a_{63} E_3) \|^2 = L_6^2.
\end{array}
\right.
\label{eq:stewart-rods}
\end{equation}

One may set $E_1 = (1,0,0), E_2 = (0,1,0), E_3 = (0,0,1)$, and then express $n$, $e_1$, and $e_2$ as linear combinations of the base frame vectors $E_1, E_2, E_3$, with coordinates $n_j = n \cdot E_j$ and $e_{ij} = e_i \cdot E_j$. The vector $e_3$ is eliminated by substituting $e_3 = e_1 \times e_2$.
In this formulation we obtain the following additional constraints:
\begin{equation}
\left\{
\begin{array}{l}
n_1^2 + n_2^2 + n_3^2 = 1, \\
e_{11}^2 + e_{12}^2 + e_{13}^2 = 1, \\
e_{21}^2 + e_{22}^2 + e_{23}^2 = 1, \\
e_{11} e_{21} + e_{12} e_{22} + e_{13} e_{23} = 0.
\end{array}
\right.
\label{eq:stewart-ortho}
\end{equation}
Combing \eqref{eq:stewart-rods} and \eqref{eq:stewart-ortho}, we have a polynomial system of $9$ equations. We treat
\[
\begin{aligned}
(&L_2, L_3, L_4, L_5, L_6, a_{21}, b_{21}, a_{31}, a_{32}, b_{31}, b_{32}, 
a_{41}, a_{42}, a_{43}, b_{41}, b_{42}, b_{43}, \\
&a_{51}, a_{52}, a_{53}, b_{51}, b_{52}, b_{53}, 
a_{61}, a_{62}, a_{63}, b_{61}, b_{62}, b_{63} )
\end{aligned}
\]
as geometric parameters and
\[
(n_1, n_2, n_3, e_{11}, e_{12}, e_{13}, e_{21}, e_{22}, e_{23})
\]
as variables.
This yields a system of $9$ equations in $9$ unknowns.
% {\color{red} What happened to $e_1$, $e_2$ and $e_3$? Provide the complete system of equations here, also specify where are parameters and which are variables. One can also tell the meaning of each subsystem.}

When generating random instances, we assign parameter values as follows. 
Here, $L_{1}$ is already fixed as $1$ by assumption. 
For $i = 2, \ldots, 6$, the rod lengths $L_{i}$ are uniformly sampled from $[0.5, 2]$. 
The coordinates $a_{21}$ and $b_{21}$ are uniformly sampled from $[0, 2]$, 
and all remaining coordinates are uniformly sampled from $[0, 1.5]$. 
These bounds were chosen to cover a special parameter configuration~\citet{dietmaier1998stewart} in which the system admits $40$ real solutions, 
although this specific configuration was not observed in any of our $10{,}000$ randomly generated instances.

%{\color{blue}
IbexSolve succeeds to solve all the $10{,}000$ generated instances, as well as the instance with $40$ real solutions reported in~\citet{dietmaier1998stewart}, within $1{,}000$ seconds, producing certified results for each instance.
Table~\ref{tab:rootcount-stewart} summarizes the distribution of root counts across these $10{,}001$ instances.
%}

\begin{table}[bht]
\centering
\small
\setlength{\tabcolsep}{6pt}
\renewcommand{\arraystretch}{1.15}
\caption{Distribution of exact root counts.}
\label{tab:rootcount-stewart}
\begin{tabular}{l|rrrrrrrrr}
\hline
Root count & 0 & 2 & 4 & 6 & 8 & 10 & 12 & 14 & 40 \\
\hline
Instances  & 7521 & 1365 & 658 & 303 & 126 & 23 & 3 & 1 & 1 \\
\hline
\end{tabular}
\end{table}

\subsubsection{Kuramoto model}\label{subsec:kuramoto}
The Kuramoto model is a classic model which describes a system of coupled oscillators with different natural frequencies, capturing how synchronization emerges from their interactions. 
The general Kuramoto model is given by: 
\begin{equation}
\frac{d\theta_i}{dt}
= \omega_i
- \frac{1}{N} \sum_{j=1}^{N} K_{i,j} \sin(\theta_i - \theta_j),
\quad \text{for } i = 1, \ldots, N,
\label{eq:kuramoto}
\end{equation}
where $N$ is the number of oscillators, $K_{i,j}$ is the coupling strength between the $i$-th and $j$-th oscillators.
Because the system is invariant under a global phase shift, it is common to fix a reference angle, e.g., set $\theta_N = 0$, and then drop the corresponding equation $\frac{d\theta_N}{dt} = 0$.
Among several equivalent formulations of the equilibrium equations~\citep{chen2019three}, we adopt a polynomial formulation for its simplicity.
Specifically, we set $K_{i,j} = 1$, corresponding to assuming uniform coupling, which is a common simplification. Then
for $i=1,\ldots,{N-1}$,
\[
\left\{
\begin{aligned}
& \omega_i - \frac{1}{N}\sum_{j=1}^N (s_i c_j - s_j c_i) = 0,\\
& c_i^2 + s_i^2 - 1 = 0,
\end{aligned}
\right.
\]
with $c_N = 1, s_N = 0$. 
This is a polynomial system with $2(N-1)$ equations and $2(N-1)$ unknowns.
It can be solved by various approaches~\citep{coss2018locating}; in this work, we focus on the subdivision method.
When generating the examples, we need to fix the parameters $\omega_i$ for $i=1,\ldots, N-1$.
To reflect practical scenarios, we sample each $\omega_i$ from a zero-mean Gaussian distribution with standard deviation $0.3$, i.e., $\omega_i \sim \mathcal{N}(0,\,0.3^2)$.
Under this setting, we generated $5{,}000$ random instances with $N=6$. 
Each instance is a square polynomial system with $10$ equations in $10$ variables.
The intervals of interest for $c_i$ and $s_i$ are set to $[-1,1]$.
IbexSolve successfully solves all 10,000 systems and returns certified results for each of them. Table~\ref{tab:rootcount-kuramoto} reports the distribution of the root counts.
% \begin{table}[hbt]
% \centering
% \small
% \setlength{\tabcolsep}{6pt}
% \begin{tabular}{r|rrrrrrrrrrrrr}
% \hline
% Root count & 10 & 12 & 16 & 18 & 20 & 22 & 24 & 26 & 28 & 29 & 30 & 31 & 32 \\
% \hline
% Instances    & 46 & 4  & 15 & 3  & 10 & 809& 65 & 143& 464 ({\bf 463}) & 2 (0) & 4176 (4179) & 2 (0) & 4261 (4263)\\
% \hline
% \end{tabular}
% \caption{Distribution of exact root counts.}
% \label{tab:rootcount-kuramoto}
% \end{table}
\begin{table}[hbt]
\centering
\small
\setlength{\tabcolsep}{6pt}
\caption{Distribution of exact root counts.}
\label{tab:rootcount-kuramoto}
\begin{tabular}{l|rrrrrrrrrrrrr}
\hline
Root count & 10 & 12 & 16 & 18 & 20 & 22 & 24 & 26 & 289 & 30 & 32 \\
\hline
Instances    & 46 & 4  & 15 & 3  & 10 & 809& 65 & 143 & 464  & 4180 & 4261 \\
\hline
\end{tabular}
\end{table}
%{\color{blue}

Since the instances are relatively small and fall within the capability of symbolic methods, we use Maple’s RootFinding package to cross-check the number of real roots. RootFinding returns results that are consistent across the instances reported in Table~\ref{tab:rootcount-kuramoto}.

However, we identified four additional instances (not listed in Table~\ref{tab:rootcount-kuramoto}) for which IbexSolve reports one fewer certified solution than RootFinding, while no suspect/unknown region is reported. This reveals an inconsistency between the two solvers: the certified enclosures returned by IbexSolve appear incomplete in these cases.

To investigate this discrepancy, we conducted a set of controlled experiments by modifying IbexSolve’s default contractor. In particular, when we disable the linear relaxation component, IbexSolve recovers the missing solution and its certified solution count becomes consistent with Maple for all four instances.

A plausible explanation is that the linear relaxation step relies on floating-point LP computations. While such relaxations can strengthen pruning in practice, the LP component relies on floating-point solvers, and floating-point roundoff can, in rare cases, lead to numerically fragile pruning. Similar concerns about the interaction between floating-point linear programming and rigorous interval reasoning have been pointed out in the literature (see, e.g.,~\cite{trombettoni2011inner}). The above four instances therefore serve as concrete evidence that, in rare cases, enabling LP-based linear relaxation may lead to overly aggressive pruning and the loss of a certified solution enclosure.

%}
% {\color{red}
% TODO: There are 5 examples are found going wrong by using symbolic solvers to certify.
% If forbidding the use of LP-relaxation (calling external LP solvers). 
% }

\subsubsection{Flash unit}
A flash unit is a separation device, used for generating vapor–liquid equilibrium streams by partially vaporizing a liquid mixture, and is commonly found in chemical and petrochemical industries.
We consider the flash unit model from~\citet{bublitz2023thesis}, which describes the partial vaporization of an ethanol–water mixture.
In this model, the liquid and vapor phases are assumed to be perfectly mixed and in thermodynamic equilibrium, and the unit temperature is maintained constant by heat supply.
The feed conditions, pressure drop, and operating temperature are specified, while the unknowns are the phase compositions, flow rates, and heat duty.
The constraints in this problem capture multiple balances, including phase equilibrium, material and energy conservation, thermodynamic property relations, and geometric conditions of the flash unit.
This problem involves 28 equations; for further details, please refer to~\citet{bublitz2023thesis}.
The full steady-state model consists of phase equilibrium 
(\ref{eq:flash-pressure}), 
activity-coefficient relations (\ref{eq:flash-activity}),
material and energy balances (\ref{eq:flash-balances}), 
thermodynamic volume relations (\ref{eq:flash-thermo-volume}), 
and geometric constraints (\ref{eq:flash-geometry}).

\begin{equation}
\left\{
\begin{aligned}
& K_i x_i - y_i = 0~(i = 1,2),  y_1 + y_2 = 1,  x_1 + x_2 = 1,  x^{F}_1 + x^{F}_2 = 1 \\
& h^{F} - \big( x^{F}_1 h^{F}_1 + x^{F}_2 h^{F}_2 +  h^{EF} \big) = 0, \\
& h^{L} - \big( x_1 h^{L}_1 + x_2 h^{L}_2 +  h^{EL} \big) = 0, \\
& h^{V} - \big( y_1 h^{V}_1 + y_2 h^{V}_2 +  h^{EV} \big) = 0,\\
& \Delta p - (p^{F} - p) = 0, K_i - \gamma_i\, p^{\mathrm{sat}}_i/p = 0 \quad (i = 1,2).
\end{aligned}
\right.
\label{eq:flash-pressure}
\end{equation}

\begin{equation}
\left\{
\begin{aligned}
& \gamma_i
- \frac{\exp\!\Big[
    (1 - x_i)\Big(
      \frac{\alpha_i}{x_i + \alpha_i (1 - x_i)}
      -
      \frac{\alpha_j}{\alpha_j x_i + (1 - x_i)}
    \Big)
  \Big]}{x_i + \alpha_i (1 - x_i)} = 0~(i = 1,2; j = 3-i),
\\
& \alpha_i
- \frac{(v_1 + v_2) - v_i}{v_i}
  \exp\!\left(
    -\frac{\lambda_i}{T}
  \right)
= 0~(i = 1,2).
\end{aligned}
\right.
\label{eq:flash-activity}
\end{equation}

\begin{equation}
\begin{aligned}
& F^{F} h^{F}
- F^{V} h^{V}
- F^{L} h^{L}
+ Q
= 0,
 F^{F} x^{F}_i
- F^{V} y_i
- F^{L} x_i
= 0~(i = 1,2).
\end{aligned}
\label{eq:flash-balances}
\end{equation}

\begin{equation}
\left\{
\begin{aligned}
& U
- n^{L} \Big(
    h^{L}
    - p \,\big(
      10^{5}\,
      ( x_1 v_1 + x_2 v_2 + v^{EL} )
    \big)
  \Big)
- n^{V} \Big(
    h^{V}
    - R \, (T\, z^{V})
  \Big)
= 0,
\\
& V - (V^{L} + V^{V}) = 0,
\\
& n_i - \big( x_i\,n^{L} + y_i\,n^{V} \big) = 0,
\quad i = 1,2,
\\
& V^{L}
- \big( (v_1 x_1 + v_2 x_2 + v^{EL})\, n^{L} \big)
= 0,
\\
& V^{V}
- \frac{
    n^{V}\, R\, (T\, z^{V})
  }{
    p \cdot 10^{5}
  }
= 0.
\end{aligned}
\right.
\label{eq:flash-thermo-volume}
\end{equation}

\begin{equation}
\begin{aligned}
& H^{L}
- \frac{4 V^{L}}{\pi D^{2}}
= 0,
A - \frac{\pi D^{2}}{4} = 0, V - A H = 0,
r - \frac{D}{H} = 0.
\end{aligned}
\label{eq:flash-geometry}
\end{equation}

The variables in this system are
$
K_1, K_2,
x_1, x_2, x^{F}_2,
y_1, y_2,
\gamma_1, \gamma_2,
\alpha_1, \alpha_2,
p,
h^{F}, h^{L}, h^{V},
F^{L}, \\
F^{V},
Q,
n_1, n_2,
n^{L}, n^{V},
U,
V^{L}, V^{V}, V,
A,
r
.
$
Here $K_i$ are vapor–liquid equilibrium ratios, $x_i$ and $y_i$ are the liquid- and vapor-phase mole fractions of component $i$, and $x^{F}_2$ is the feed composition of component $2$. The quantities $\gamma_i$ and $\alpha_i$ are, respectively, the activity coefficients and their temperature-dependent parameters. The scalar $p$ is the operating pressure. The symbols $h^{F}, h^{L}, h^{V}$ denote the specific enthalpies of the feed, liquid phase, and vapor phase; $F^{L}$ and $F^{V}$ are the liquid and vapor outlet flow rates; and $Q$ is the heat duty. The variables $n_i$ are the holdup amounts of component $i$ in the vessel, while $n^{L}$ and $n^{V}$ are the total liquid and vapor holdups. The quantity $U$ is the total internal energy. The variables $V^{L}$, $V^{V}$, and $V$ are the liquid, vapor, and total volumes. Finally, $A$ is the vessel cross-sectional area and $r$ is the diameter-to-height ratio.

The rest of the symbols in the system are treated as parameters.
To generate instance, we fix the following parameters relating to the operating conditions, thermophysical constants, or geometric data with the data in~\citet{bublitz2023thesis}: the flash temperature $T=353.15\,\mathrm{K}$ and feed temperature $T^{F}=353.15\,\mathrm{K}$; the enthalpy offset terms $h^{EF}$, $h^{EL}$, $h^{EV}$ (taken as $0$ in the present setting); the vessel diameter $D=0.16$, characteristic height $H=0.5$, and reference liquid height $H^{L}=0.25$; the excess liquid volume contribution $v^{EL}=0$; and the vapor compressibility factor $z^{V}=1.0$. Thermodynamic correlations are encoded by per-component parameters for enthalpy and saturation pressure. For each component $i\in\{1,2\}$, the liquid-, vapor-, and feed-phase enthalpies $h^{L}_i$, $h^{V}_i$, $h^{F}_i$ are evaluated from expressions with coefficients given numerically in the implementation (see p.187 in \citet{bublitz2023thesis} for details), as well as the saturation pressures $p_i^{\mathrm{sat}}$. The parameters $(\lambda_1,\lambda_2)=(95.68,506.7)$ are empirical constants and the liquid molar volumes $(v_1,v_2)=(5.869\times10^{-5},\,1.807\times10^{-5})$, all in consistent units. Finally, the gas constant is $R=8.314$, which enters the vapor holdup and volume relations.

Apart from the fixed parameters, several operating inputs are treated as
user-specified: the pressure drop $\Delta p$,
the feed pressure $p^{F}$, the feed composition $x^{F}_1$ of component 1, and the feed flow rate $F^{F}$.
To generate benchmark instances, we systematically vary these
user-specified parameters over predefined ranges with fixed step sizes.
By combining these variations on a grid, we obtain a total of
$10{,}000$ problem instances.
The choice of parameter ranges is motivated by the values reported in
\citet{bublitz2023thesis}, and extended to cover practically relevant
operating conditions.
%{\color{blue}
The region of interest for each variable is set to $[-10^9,10^9]$ as in \citet{bublitz2023thesis}.
%}

%{\color{blue}
IbexSolve successfully produced outputs for all 10,000 instances. However, none of these outputs are fully certified: every instance contains at least one unknown box. In addition to the unknown boxes, $3436$ of the $10000$ instances also contain certified solution boxes, $1956$ instances produce one solution box, $1290$ produce two and $190$ produce three.
The prevalence of unknown boxes across all instances appears to stem mainly from two factors. First, the initial search domains are extremely wide. Second, the systems are high-dimensional and strongly nonlinear, combining multivariate polynomials with exponential terms and rational expressions. 
These features make rigorous interval-based certification much more difficult, especially when wide domains allow denominators to approach zero. Empirically, once the initial domains are tightened to physically meaningful ranges, the frequency of unknown boxes decreases substantially.
%}

% {\color{red} There are unknown boxes for every instance. default box is 10-9~10-9. 
% Explain why: intitial boxes are too large, strongly nonlinear, Jacobian is close to be deficient. 
% }

\subsubsection{Initial Orbit determination}
Initial Orbit Determination (IOD) estimates a body’s Keplerian orbit from minimal observations—often angles-only (lines of sight), with no prior information. Geometrically, this reduces to finding a conic with a specified focus that meets the observed lines of sight. Here we consider the following problem: find a conic in $\mathbb{R}^3$  with one focus at the origin that is incident to five given, generically positioned lines. This admits a formulation as a square system of nonlinear equations. As in other practical settings, the problem has multiple equivalent algebraic reformulations~\citep{duff2022geometric,leykin2025projective}; here we adopt the formulation of~\citet{leykin2025projective}.
\begin{figure}[h!]
    \centering
    \includegraphics[width=0.5\textwidth]{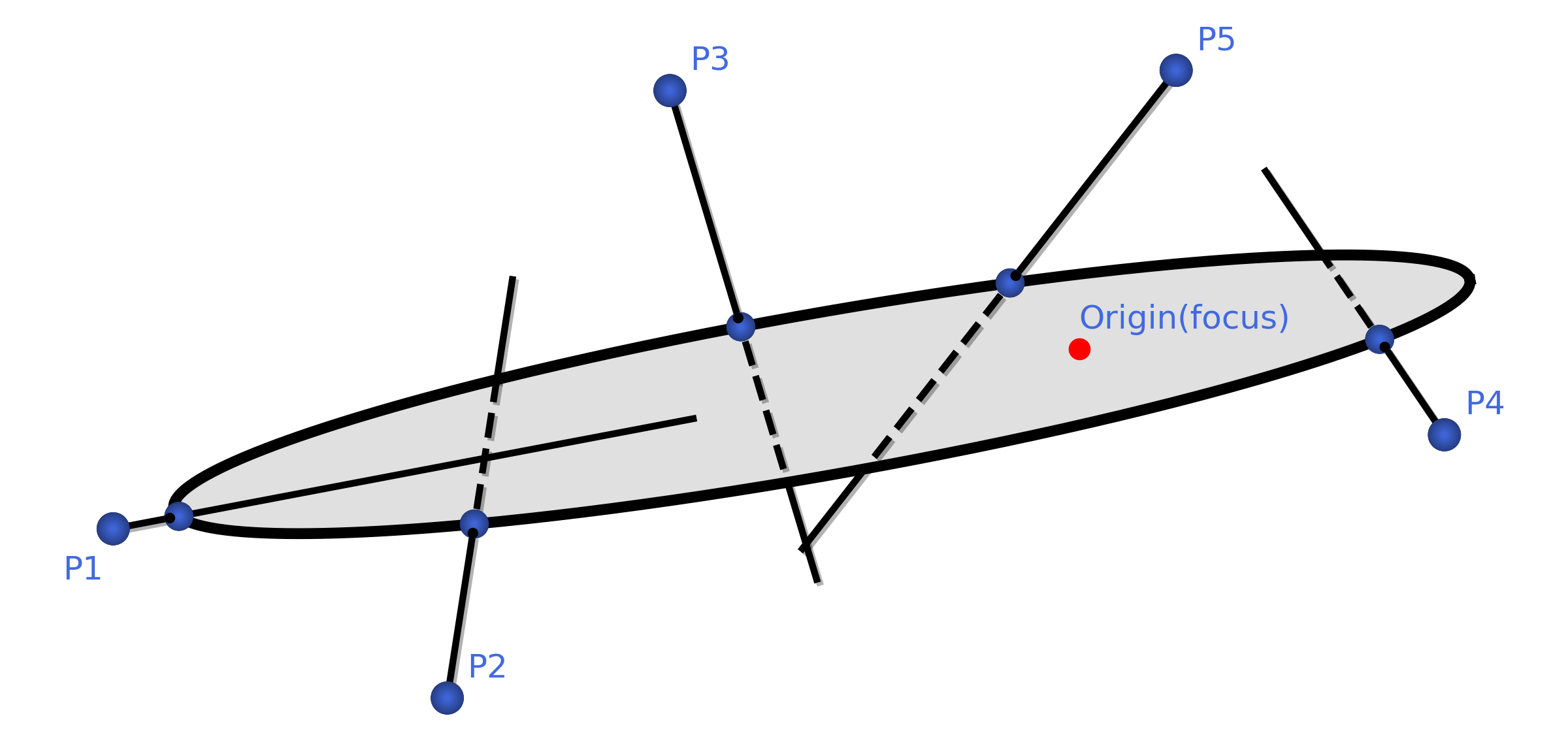} 
    \caption{Geometric setup: the points $P_i$ mark observer positions, and the lines depict the observed lines of sight.}
    \label{fig:orbit}
\end{figure}

Here we briefly describe the known data, variables and the constraints.
The positions of the five observers are given in the known $3\times5$ matrix $P$, and their lines of sight (unit direction vectors) are given in the known $3\times5$ matrix $U$.
We take $w=(w_1,w_2,w_3)\in \mathbb{R}^3$ as the unit normal to the orbital plane and, by symmetry, assume $w_1\ge 0$.
This introduces three variables $w_1,w_2,w_3$ with the region of interest $w_1\in [0,1],  w_2, w_3 \in [-1,1]$, 
and a constraint $\|w\|^2=1$.
To complete a local orthonormal frame on the orbital plane, we need to add two unit vectors $v, v'\in \RR^3$.
Following~\citet{leykin2025projective},
let $e_1=(1,0,0)$, set
 $v' = \lambda'(w\times e_1)$, with $\lambda' = \frac{1}{\|w\times e_1\|}$, and define $v = v'\times w$.
Here $\lambda'$ is introduced as a variable,
with region of interest $[1,\infty)$; the corresponding constraint is $\lambda' = \frac{1}{\|w\times e_1\|}$.

We introduce scalar variables $\rho_i$ for the intersections of each line of sight with the orbital plane. Let $p_i$ and $u_i$ be the $i$-th columns of $P$ and $U$, respectively, and define
$$
r_i = p_i + \rho_i u_i,\quad i = 1,\ldots,5
$$
subject to the coplanarity conditions
$r_i \cdot w = 0$ for $i = 1,\ldots,5$
since vectors in the orbital plane are orthogonal to the normal $w$. This introduces five variables $\{\rho_i\}_{i=1}^5$ and five linear constraints.

In the local orthonormal frame, the conic (after normalization) takes the form $a x^{2}+b x y+c y^{2}+d x+e y+1=0$, with five variables $a,b,c,d,e$. Plugging the intersection points $r_i$ (expressed in this frame) into the conic equation yields five constraints, one for each $i=1,\ldots,5$.

The requirement that the focus lie at the origin imposes two additional constraints~\citep{semple1998algebraic}:
\[
e^2 - d^2 + 4a - 4c = 0, \quad
de -2b =0.
\]

Together, these equations yield a square polynomial system in $14$ variables as follows:
\begin{equation}
\left\{
\begin{array}{l}
\|w\|^2 = 1, \\
\lambda'^2 \cdot\|w\times e_1\|^2 = 1, \\
w\cdot (p_i + \rho_i u_i) = 0, \quad i= 1,\ldots,5, \\
a\, x_i^{2}+b\, x_i y_i+c\, y_i^{2}+d\, x_i+e\, y_i+1=0, \quad i = 1,\ldots,5, \\
e^2 - d^2 + 4a - 4c = 0, \\
de -2b =0,
\end{array}
\right.
\label{eq:orbit}
\end{equation}
where $w=(w_1,w_2,w_3)$, $e_1= (1,0,0)$, $p_i$ and $u_i$ are respectively the $i$-th columns of matrices $P$ and $U$, the coordinates $x_i = (p_i + \rho_i u_i) \cdot v$, $y_i = (p_i + \rho_i u_i) \cdot v'$ with  $v' = \lambda'(w\times e_1)$, and $v = v'\times w$.
In this system, the matrix $P$ and $U$ are given data, and the unknowns are $w_1, w_2, w_3, \lambda', \rho_i (i=1,\ldots,5), a, b,c,d,e$. 

To generate instances, we construct the position matrix $P\in\mathbb{R}^{3\times5}$ by drawing $p_i\in\mathbb{R}^3$ independently and identically distributed from the uniform box $[-10,10]^3$ and stacking them as the columns of $P$.
The direction matrix $U\in\mathbb{R}^{3\times5}$ is formed by sampling vector $z_i\in\RR^3$ independently from standard normal distribution and normalizing $u_i = z_i / \lVert z_i\rVert$, and stacking the unit vectors $u_i$ as the columns of $U$.
Using this procedure, we generated $2{,}000$ instances.

%{\color{blue}
These instances are fairly challenging: when solved with IbexSolve, most of them time out. To accelerate the search, we augment the original square system with an additional inequality for each instance, $b^2-4ac<0$,which restricts the trajectory to be elliptical. 
%This restriction matches typical orbital behavior in practice, such as the trajectories of artificial satellites and planets.
Under this setting, IbexSolve solved $1{,}723$ instances with certified results, while the rest $277$ instances timed out.
Table~\ref{tab:rootcount-orbit} reports the distribution of the root counts over the $1{,}723$ instances.
%}
\begin{table}[bht]
\centering
\small
\setlength{\tabcolsep}{4pt}
\renewcommand{\arraystretch}{1.15}
\caption{Distribution of root counts.}
\label{tab:rootcount-orbit}
\begin{tabular}{c|rrrrrrrrrrrrr}
\hline
Root count & 0 & 1 & 2 & 3 & 4 & 5 & 6 & 7 & 8 & 9 & 10 & 11 & 13 \\
\hline
Instances  & 118 & 182 & 315 & 275 & 294 & 222 & 145 & 87 & 45 & 26 & 11 & 2 & 1 \\
\hline
\end{tabular}
\end{table}

%{\color{red} an inequality is added to the system.}

%\subsection{Key findings}

\section{Application of the dataset}

\subsection{As  a benchmark dataset}\label{subsec:application1}
    To compare the performance of different solvers and to collect solution information for the benchmarks, we ran these examples on IbexSolve, RealPaver, and Maple. 
    We will reports and compare the results across these solvers.
    
    We first consider the two subdivision-based solvers, IbexSolve and RealPaver. The solvers, although sharing a common subdivision framework, differs in various aspects, such as the contraction strategy, bisection strategy, node selection strategy and other implementation details, which leads to differences in efficiency and, in some cases, in the outcomes.
    IbexSolve uses depth-first search by default. Its default contractor is a composition of several techniques, including hull consistency technique, the Hansen–Sengupta operator, the ACID contractor, and linear-relaxation–based contractors.
    The default bisection strategy is the Relative Smear Sum (RSS) strategy.
    For zero-dimensional square systems, IbexSolve subdivides the entire region of interest and outputs solution boxes, which are verified to contain a unique solution, and suspect boxes, which can neither be discarded nor certified for uniqueness.
    RealPaver employs a composite contractor that integrates hull and box consistency techniques, the Hansen–Sengupta operator, and 3B-consistency methods.
    By default, RealPaver performs depth-first search and uses a bisection policy that combines round-robin with a max-narrow rule (selecting the variable whose Jacobian column has the largest absolute sum; if the Jacobian is unavailable, it falls back to round-robin). However, in our experiments, this strategy caused 18 instances to produce outputs exceeding 1 GB when no cap on the number of boxes was set, indicating an output blow-up under this setting. We therefore switched to paving mode, which adopts breadth-first search with a largest-first bisection policy. 
%    On the other hand, 6 instances in paving mode were terminated by the OOM-killer (out-of-memory). % only on my computer
%    Overall, at least in our environment, RealPaver proved less robust than IbexSolve in terms of memory usage and output handling.
    Apart from these outliers, the overall performance of paving mode and the default mode is comparable. On balance, we report results using paving mode. Unless otherwise noted, all RealPaver results below are for paving mode.
    RealPaver outputs all boxes that meet the prescribed precision and cannot be pruned by its contractors; unlike IbexSolve, in most cases, it does not certify and label boxes as containing a unique solution, even though verification-style operators (e.g., Hansen–Sengupta) may be used internally as contractors. 
    For fairness, we fix the bisection precision to the same value for both solvers, setting it to $10^{-6}$, and impose a timeout of 1000 seconds.
    
%    Among the XX instances, X runs failed in RealPaver due to unrecognized syntax, X runs were terminated by the OOM-killer, and X runs failed in IbexSolve with a segmentation fault (possibly a bug).
    Among the 581 instances, 1 runs failed in RealPaver due to unrecognized syntax, and 2 runs failed in IbexSolve with a segmentation fault (possibly a bug).
    Of the remaining instances, 459 were solved by IbexSolve within 1000 seconds; among these, 393 produced only certified boxes (i.e., no suspect regions). RealPaver solved 385 instances within the same time limit.

\begin{table}[ht]
\centering
\caption{Numbers of solved examples in different timing intervals.}
\label{tab:count}
\begin{tabular}{l|c|c|c|c|c}
\hline
Timing (s) & $\le$ 1 & 1-10 & 10-100 & 100-1000 & Timeout ($>$1000) \\
\hline
IbexSolve & 288 & 83 & 47 & 41 & 120 \\
RealPaver & 268 & 49 & 37 & 31 & 195 \\
\hline
\end{tabular}
\end{table}    
    
% \begin{table}[ht]
% \centering
% \caption{Distribution of timing of Ibexsolve (in seconds)}
% \begin{tabular}{l|c|c|c|c|c}
% \hline
% Timing & $\le$ 1 & 1-10 & 10-100 & 100-1000 & Timeout ($>$1000) \\
% \hline
% Count & 318 & 87 & 47 & 42 & 125 \\
% \hline
% \end{tabular}
% \end{table}
    
% \begin{table}[ht]
% \centering
% \caption{Distribution of timing of RealPaver (in seconds)}
% \begin{tabular}{l|c|c|c|c|c}
% \hline
% Timing & $\le$ 1 & 1-10 & 10-100 & 100-1000 & Timeout ($>$1000) \\
% \hline
% Count & 294 & 53 & 42 & 32 & 201 \\
% \hline
% \end{tabular}
% \end{table}

\begin{figure}[h!]
    \centering
    \includegraphics[width=0.8\textwidth]{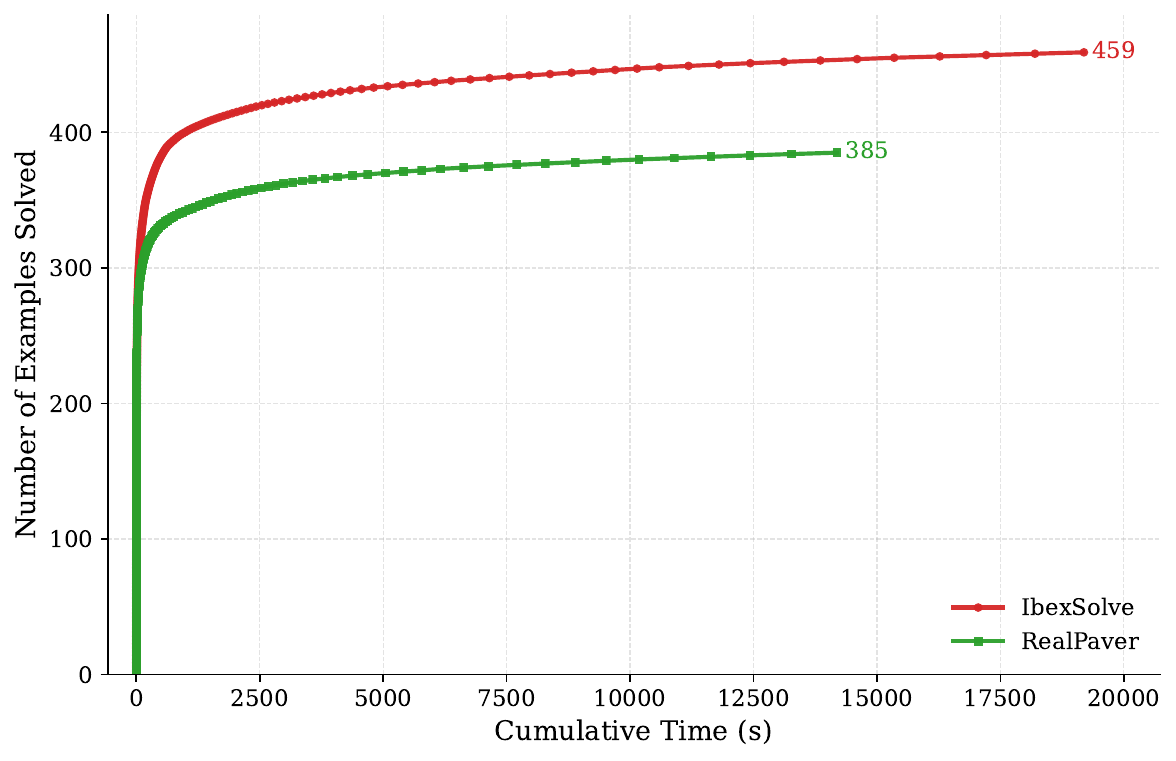}
    \caption{Cumulative runtime comparison: IbexSolve and RealPaver. }
    \label{fig:cumul-runtime}
    \end{figure}

Table~\ref{tab:count} counts 
the number of solved examples in different timing intervals.
    Figure~\ref{fig:cumul-runtime} further reports the cumulative runtime comparison between IbexSolve and RealPaver, 
    where instances that timed out are excluded.
    From both Table~\ref{tab:count} and Figure~\ref{fig:cumul-runtime}, we observe that, statistically, IbexSolve is substantially faster than RealPaver.
%     {\color{red}
%     TODO: Stewgou should be removed from and fonts should be made larger in Figure~\ref{fig:cumul-runtime}?
% }    
    
    % This is made clearer in Figure~\ref{fig:runtime}, which plots a solve-count curve: the $x$-axis is wall-clock time $t$, and the $y$-axis reports the count of instances with runtime $\le t$. Note that time is not accumulated.
    % \begin{figure}[h!]
    % \centering
    % \includegraphics[width=0.5\textwidth]{task7/old/dataset_compare_chart.png}
    % \caption{Solve-count curves for IbexSolve and RealPaver}
    % \label{fig:runtime}
    % \end{figure}
    
    Across the 581 instances, IbexSolve exclusively solved 102/581 (17.6\%) and RealPaver exclusively solved 28/581 (4.8\%)—i.e., solved by one solver while the other timed out or errored. Both solved 357/581 (61.4\%); among these, excluding the 241 easy cases solved by both within 1 second, IbexSolve was faster in 85 cases (74.6\%) and RealPaver in 29 (25.4\%).

\newcommand{\legendbox}[1]{\tikz{\path[fill=#1,draw=#1] (0,0) rectangle (0.25,0.25);} }

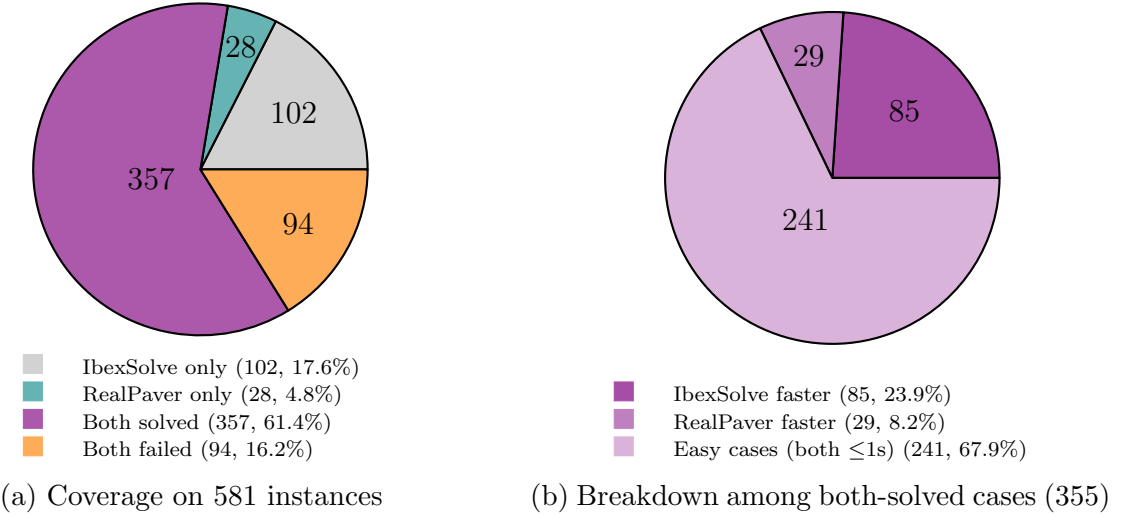
\begin{figure}[ht]
  \centering

  % ---------- Panel A: 覆盖率（583） ----------
  \begin{subfigure}{0.48\linewidth}
    \centering
    \begin{tikzpicture}
      \pie[
        sum=auto,
        radius=2.2,
        color={gray!35, teal!60, violet!65, orange!65}
      ]{
        102/,   % IbexSolve only
        28/,    % RealPaver only
        357/,   % Both solved
        94/     % Both failed
      }
    \end{tikzpicture}

    % 手工图例（对齐版）
    \vspace{4pt}\scriptsize
    \begin{tabular}{@{}l l@{}}
      \legendbox{gray!35}  & IbexSolve only (102, 17.6\%) \\
      \legendbox{teal!60}  & RealPaver only (28, 4.8\%) \\
      \legendbox{violet!65} & Both solved (357, 61.4\%) \\
      \legendbox{orange!65}& Both failed (94, 16.2\%)
    \end{tabular}

    \caption{Coverage on 581 instances}
  \end{subfigure}
  \hfill
  % ---------- Panel B: 双方均解出的 360 个中的分布（含 easy） ----------
  \begin{subfigure}{0.48\linewidth}
    \centering
    \begin{tikzpicture}
      % 先画与 Panel A 中 “Both solved” 一致色系的底圈（绿色）
      \fill[violet!20] (0,0) circle (2.2cm);
      % 再叠加细分饼图（绿色同色系深浅）
      \pie[
        sum=auto,
        radius=2.2,
        color={violet!70, violet!50, violet!30}
      ]{
        85/,    % IbexSolve faster (non-easy)
        29/,    % RealPaver faster (non-easy)
        241/    % Easy cases (both ≤1s)
      }
    \end{tikzpicture}

    \vspace{4pt}\scriptsize
    \begin{tabular}{@{}l l@{}}
    \legendbox{violet!70} & IbexSolve faster (85, 23.9\%) \\
    \legendbox{violet!50} & RealPaver faster (29, 8.2\%) \\
    \legendbox{violet!30} & Easy cases (both $\le$1s) (241, 67.9\%) \\
    \end{tabular}

    \caption{Breakdown among both-solved cases (355)}
  \end{subfigure}

  \caption{IbexSolve vs.\ RealPaver: comparison of timing.}
\end{figure}

    In addition to the subdivision-based solvers, we also ran the instances in Maple using \texttt{RootFinding[Isolate]}, which is based on symbolic computation. Because symbolic methods are applicable only to polynomial systems, we restricted the Maple experiments to the polynomial subset of our benchmark.
    The output of \texttt{RootFinding[Isolate]} is a list of isolating intervals/boxes for the real solutions of the input polynomial system. The routine is designed to compute all real solutions and does not support restricting the search to a user-specified region of interest.
    Among 451 polynomial instances, Maple produced solutions for 278; the remainder either timed out, were terminated due to a kernel crash, or were inapplicable because the system defines a positive-dimensional complex variety.
    
\begin{table}[ht]
\centering
\caption{Distribution of timing of RootFinding in Maple(in seconds).}
\begin{tabular}{l|c|c|c|c|c|c}
\hline
Timing & $\le$ 1 & 1-10 & 10-100 & 100-1000 & Timeout ($>$1000) & Error\\
\hline
Count & 201 & 40 & 19 & 18 & 156 & 17 \\
\hline
\end{tabular}
\end{table}

\begin{figure}[h!]
    \centering
    \includegraphics[width=0.8\textwidth]{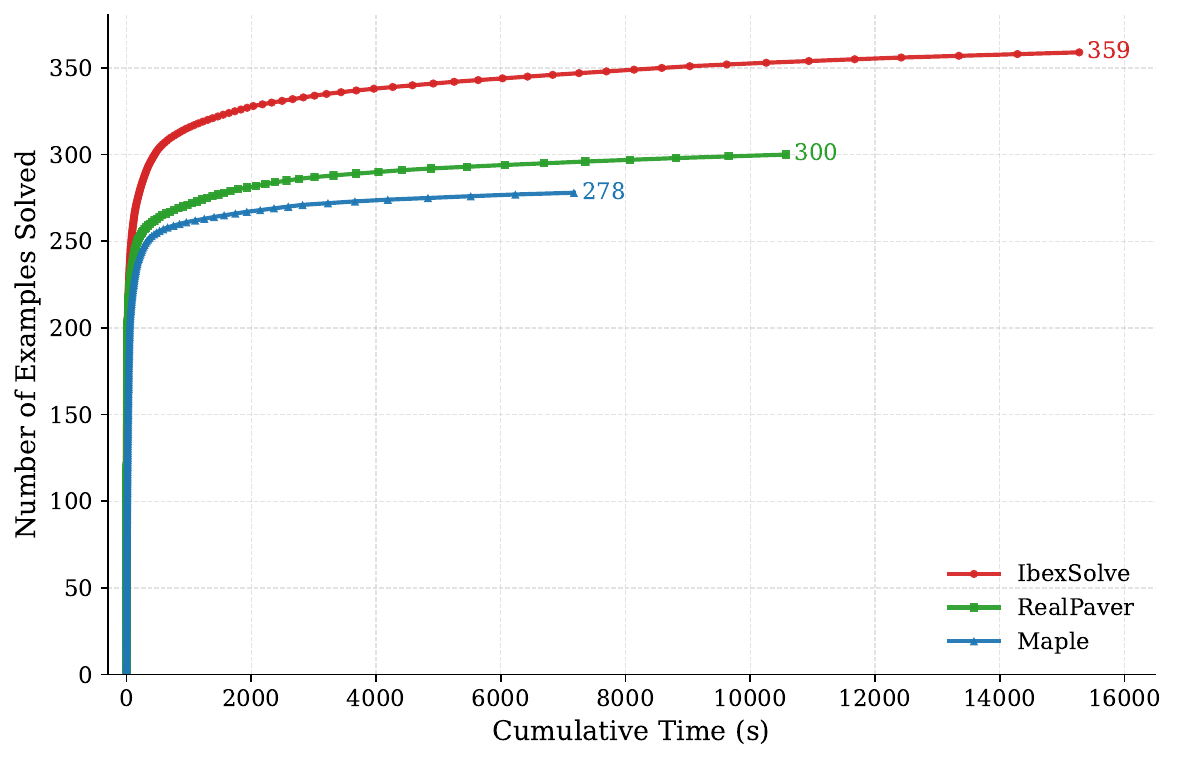}
    \caption{Cumulative runtime comparison: IbexSolve , RealPaver and Maple}
    \label{fig:cumul-runtime-three}
    \end{figure}

    Figure~\ref{fig:cumul-runtime-three} reports the cumulative runtime comparison between IbexSolve, RealPaver and Maple, restricted to the polynomial subset.
    Overall, in our experiments, IbexSolve outperforms both RealPaver and Maple’s \texttt{RootFinding[Isolate]}, and the latter two are comparable.
    On the other hand, as seen in the IbexSolve–RealPaver comparison, no single solver is uniformly faster than another across instances. 
    We therefore compare Maple’s \texttt{RootFinding[Isolate]} against IbexSolve and RealPaver separately.
Across the 451 instances, IbexSolve exclusively solved 104/451 (23.1\%) and Maple exclusively solved 23/451 (5.1\%), i.e., solved by one solver while the other timed out or errored; among the 104 instances solved only by IbexSolve, 80 were exact and 24 were non-exact. Both solved 255/451 (56.5\%), among which IbexSolve was exact in 222 cases and non-exact in 33 cases. Excluding the 146 easy cases both solved within 1 second (119 exact and 27 non-exact for IbexSolve), IbexSolve was faster in 51 cases (46.8\%; all exact) and Maple in 58 (53.2\%; 52 exact and 6 non-exact for IbexSolve).
    
\begin{figure}[ht]
  \centering
  \begin{subfigure}{0.48\linewidth}
    \centering
% preamble needs:
% \usetikzlibrary{patterns}

% preamble: \usetikzlibrary{patterns}

\begin{tikzpicture}
  % base pie (unchanged)
  \pie[
    sum=auto,
    radius=2.2,
    color={gray!35, teal!60, violet!65, orange!65}
  ]{
    104/,  % IbexSolve only
    23/,   % Maple only
    255/,  % Both solved
    69/    % Both failed
  }

  % hatch from center for IbexSolve non-exact parts (precomputed angles, total=451)
  \def\R{2.2cm}

  % IbexSolve only: 104, non-exact=24  -> hatch on last part of gray sector
  % angles: gray sector 0 -> 83.015521; hatch 63.857982 -> 83.015521
  \path[pattern=north east lines, pattern color=black!70, draw=none]
    (0,0) -- (63.857982:\R) arc[start angle=63.857982,end angle=83.015521,radius=\R] -- cycle;

  % Both solved: 255, non-exact=33 -> hatch on last part of purple sector
  % angles: both sector 101.374723 -> 304.878049; hatch 278.536585 -> 304.878049
  \path[pattern=north east lines, pattern color=black!70, draw=none]
    (0,0) -- (278.536585:\R) arc[start angle=278.536585,end angle=304.878049,radius=\R] -- cycle;
\end{tikzpicture}

    % 手工图例（对齐版）
    \vspace{4pt}\scriptsize
    \begin{tabular}{@{}l l@{}}
      \legendbox{gray!35}   & IbexSolve only (104, 23.1\%) \\
      \legendbox{teal!60}   & Maple only (23, 5.1\%) \\
      \legendbox{violet!65} & Both solved (255, 56.5\%) \\
      \legendbox{orange!65} & Both failed (69, 15.3\%)
      
    \end{tabular}

    \caption{Coverage on 451 instances}
  \end{subfigure}
  \hfill
  % ---------- Panel B: 双方均解出的 255 个中的分布（含 easy） ----------
  \begin{subfigure}{0.48\linewidth}
    \centering
% preamble: \usetikzlibrary{patterns}

\begin{tikzpicture}
  % 底圈与 Panel A 的 “Both solved” 配色一致系
  \fill[violet!20] (0,0) circle (2.2cm);

  % 细分饼图
  \pie[
    sum=auto,
    radius=2.2,
    color={violet!70, violet!50, violet!30}
  ]{
    51/,    % IbexSolve faster (non-easy)
    58/,    % Maple faster (non-easy)
    146/    % Easy cases (both ≤1s)
  }

  % ---- Hatch overlay: IbexSolve non-exact within sectors ----
  % total = 255
  % slice1 (51):   0 -> 72.000000
  % slice2 (58):  72.000000 -> 153.882353   non-exact=6 => last 8.470588 deg: 145.411765 -> 153.882353
  % slice3 (146): 153.882353 -> 360.000000  non-exact=27 => last 38.117647 deg: 321.882353 -> 360.000000
  \def\R{2.2cm}

  % Maple faster: IbexSolve non-exact = 6
  \path[pattern=north east lines, pattern color=black!70, draw=none]
    (0,0) -- (145.411765:\R)
    arc[start angle=145.411765, end angle=153.882353, radius=\R] -- cycle;

  % Easy: IbexSolve non-exact = 27
  \path[pattern=north east lines, pattern color=black!70, draw=none]
    (0,0) -- (321.882353:\R)
    arc[start angle=321.882353, end angle=360.000000, radius=\R] -- cycle;

\end{tikzpicture}

    \vspace{4pt}\scriptsize
    \begin{tabular}{@{}l l@{}}
      \legendbox{violet!70} & IbexSolve faster (51, 20.0\%) \\
      \legendbox{violet!50} & Maple faster (58, 22.7\%) \\
      \legendbox{violet!30} & Easy cases (both $\le$1s) (146, 57.3\%)
    \end{tabular}

    \caption{Breakdown among both-solved cases (255)}
  \end{subfigure}
{\scriptsize {Hatched areas mark instances with non-certified IbexSolve output}}
  \caption{IbexSolve vs.\ Maple: comparison of timing}
\end{figure}
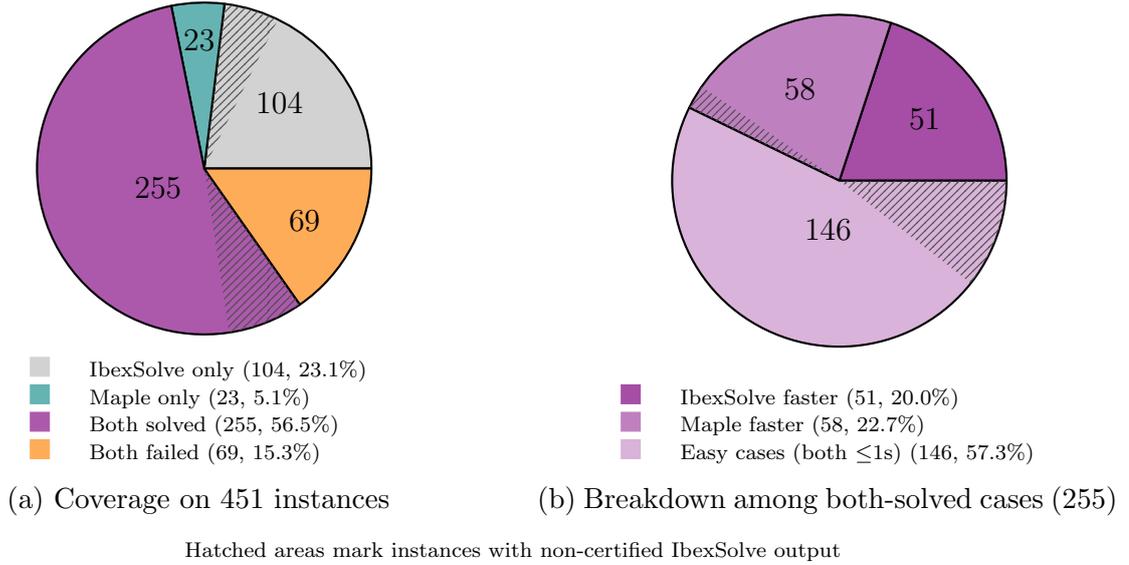

Across the 451 instances, RealPaver exclusively solved 88/451 (19.5\%) and Maple exclusively solved 66/451 (14.6\%). Both solved 212/451 (47.0\%); among these, excluding the 126 easy cases both solved within 1 second, RealPaver was faster in 38 cases (44.2\%) and Maple in 48 (55.8\%).
   
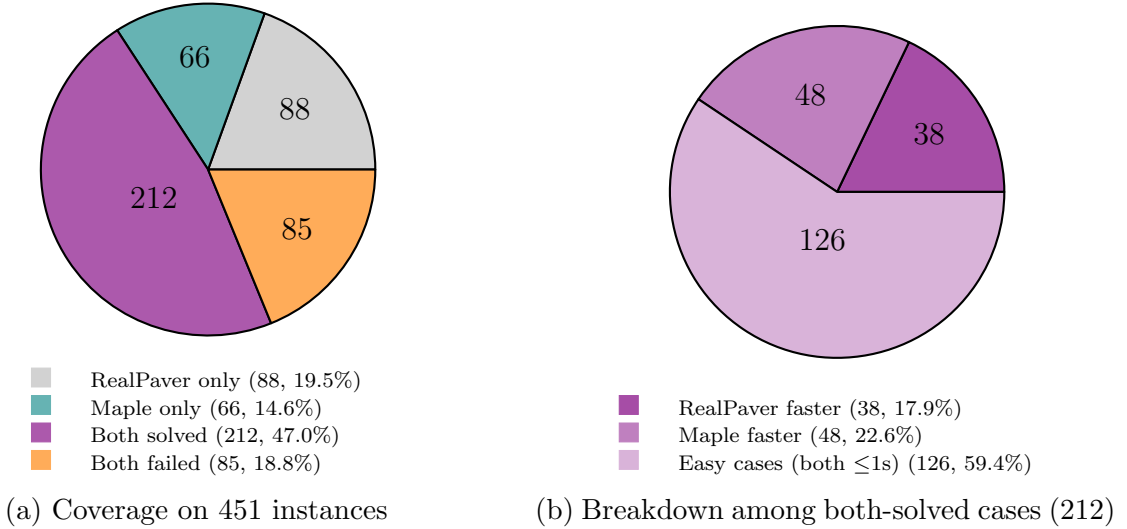
\begin{figure}[ht]
  \centering

  \begin{subfigure}{0.48\linewidth}
    \centering
    \begin{tikzpicture}
      \pie[
        sum=auto,
        radius=2.2,
        color={gray!35, teal!60, violet!65, orange!65}
      ]{
        88/,   % RealPaver only
        66/,   % Maple only
        212/,  % Both solved
        85/    % Both failed
      }
    \end{tikzpicture}

    % 手工图例（对齐版）
    \vspace{4pt}\scriptsize
    \begin{tabular}{@{}l l@{}}
      \legendbox{gray!35}   & RealPaver only (88, 19.5\%) \\
      \legendbox{teal!60}   & Maple only (66, 14.6\%) \\
      \legendbox{violet!65} & Both solved (212, 47.0\%) \\
      \legendbox{orange!65} & Both failed (85, 18.8\%)
    \end{tabular}

    \caption{Coverage on 451 instances}
  \end{subfigure}
  \hfill
  % ---------- Panel B: 双方均解出的 212 个中的分布（含 easy） ----------
  \begin{subfigure}{0.48\linewidth}
    \centering
    \begin{tikzpicture}
      % 底圈与 Panel A 的 “Both solved” 配色一致系
      \fill[violet!20] (0,0) circle (2.2cm);
      % 细分饼图
      \pie[
        sum=auto,
        radius=2.2,
        color={violet!70, violet!50, violet!30}
      ]{
        38/,    % RealPaver faster (non-easy)
        48/,    % Maple faster (non-easy)
        126/    % Easy cases (both ≤1s)
      }
    \end{tikzpicture}

    \vspace{4pt}\scriptsize
    \begin{tabular}{@{}l l@{}}
      \legendbox{violet!70} & RealPaver faster (38, 17.9\%) \\
      \legendbox{violet!50} & Maple faster (48, 22.6\%) \\
      \legendbox{violet!30} & Easy cases (both $\le$1s) (126, 59.4\%)
    \end{tabular}

    \caption{Breakdown among both-solved cases (212)}
  \end{subfigure}

  \caption{RealPaver vs.\ Maple: comparison of timing}
\end{figure}

% {\color{red}In both the left subfigures of figure 9 and 10, for the purple and grey area, maybe we should further divide them into two regions: one region is that the subdivision method agrees with the symbolic method (or certified) and the rest are those not agreed.}

    % \cored{ Todo: Maybe further check the instances on which maple is faster, due to the presence of multiple root?}
    
    To assess consistency across methods,
    we compare the solutions returned by IbexSolve and Maple. RealPaver is excluded at this stage because it does not certify roots (e.g., it does not prove uniqueness).
    From our experiments, we observe 31 instances in which Maple isolates more solutions within the region of interest than IbexSolve. However, this does not imply that IbexSolve’s search missed solutions: in every such instance, the additional Maple solutions fall inside IbexSolve’s suspect (unknown) boxes.
    Evaluating the Jacobian at these non-certified solutions yields $\det(J) \approx 0$, indicating the presence of multiple roots. Hence IbexSolve’s failure to certify uniqueness is most plausibly due to root multiplicity, which is consistent with its certification procedure. In this sense, the outputs of the two tools are mutually consistent.
    
%{\color{blue}
  From the above comparisons, IbexSolve is generally more efficient than RealPaver, and RealPaver tends to be faster than Maple. However, this ordering is only statistical: across individual instances, none of the solvers is universally faster than the others. This observation also echoes our earlier discussion on subdivision strategies.
  For the two subdivision-based tools, IbexSolve and RealPaver, IbexSolve is generally faster overall. However, for some high-dimensional families in our dataset, such as the Yamamura and Trigo series (with up to thousands of variables), RealPaver performs remarkably well, often succeeding where IbexSolve fails.
  Compared with the symbolic tool RootFinding, subdivision-based solvers tend to have an advantage on problems with larger numbers of variables but relatively well-behaved solution sets, especially when the number of solutions is small. In contrast, symbolic methods are often more suitable for lower-degree systems with intricate algebraic structure, especially when multiple roots are present.
  Overall, these results indicate that solver performance is highly instance-dependent, suggesting that a feature-driven strategy can be an effective way to improve efficiency.
%}

    By running IbexSolve on both the benchmarks we collected and the instances we generated, we observed several issues that suggest IbexSolve may contain implementation bugs or, at least, has room for improvement. Overall, we categorize these observations into four types: for two of them we have identified plausible causes, while the other two still require more careful investigation.

The first issue is related to the use of the linear relaxation technique, which can lead to results that are not fully certified. We have already discussed this point in \ref{subsec:kuramoto} and \ref{subsec:robot}, and our experiments provide concrete evidence of this issue.
The second issue concerns a few benchmarks where IbexSolve reports more solutions than Maple. A closer inspection suggests that this is largely due to the box inflation performed during IbexSolve’s final verification stage. If two neighboring boxes are both checked and a solution lies near their common boundary, both boxes can be certified as containing a (unique) solution, so the same solution gets counted twice. Fortunately, this overcounting can typically be corrected with a simple post-processing procedure.
The third issue concerns the spatial systems in \ref{subsec:robot}. For a small number of instances, the solutions produced by the trigonometric and polynomial formulations do not agree, even though the two formulations describe the same underlying problem. After ruling out the first two explanations above, the mismatch still remains, and we have not yet identified its cause.
Finally, we observed a fourth issue: IbexSolve crashes with a segmentation fault on two benchmarks, one of which is quite simple and should be easy to solve. Although these are isolated cases, they may indicate a latent bug in the implementation.

%{\color{blue}
Another observation comes from running IbexSolve on the robot systems in \ref{subsec:robot}. We found that, if we relax the bisection tolerance to $10^{-3}$
 (instead of the default $10^{-6}$
 used throughout the previous experiments), IbexSolve actually produces certified results for more instances within the same 1000-second time limit, excluding those that either time out or still contain suspect regions. This phenomenon is consistent across all four categories of systems in \ref{subsec:robot}. The result is somewhat counterintuitive, since higher precision is often expected to provide more information and thus facilitate certification. A plausible explanation is that a tighter tolerance forces substantially more bisection and propagation, leaving less budget for the final certification steps; in contrast, a slightly coarser tolerance may reduce the search overhead and allow more instances to reach a certified conclusion. This finding may shed light on practical ways to improve efficiency of solvers by choosing an appropriate bisection tolerance.
 %}

\subsection{For training machine learning models}
\begin{figure}[h!]
    \centering
    \includegraphics[width=1\textwidth]{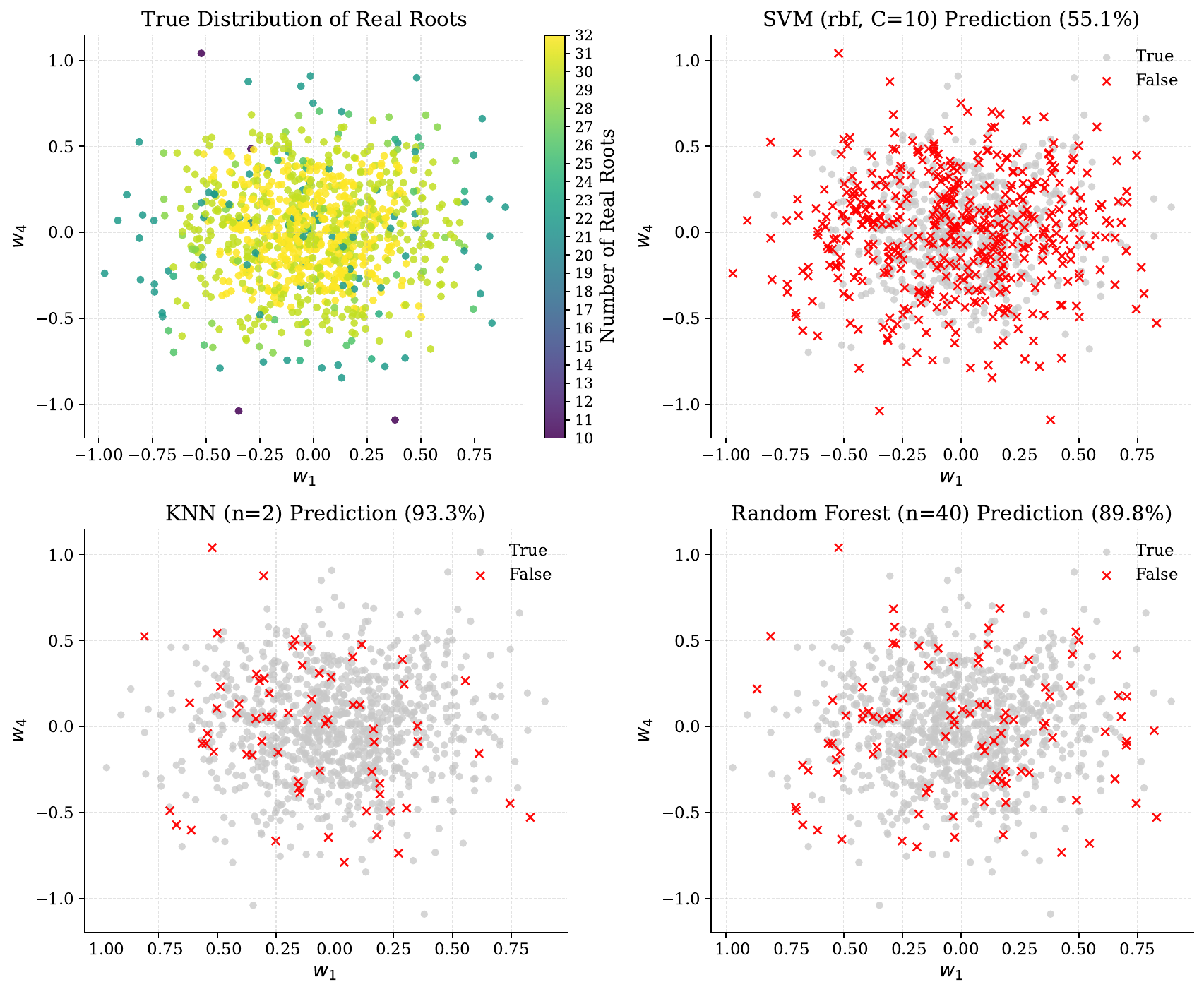} 
    \caption{The Performance Visualization of Models on Test Set.}
    \label{fig:kuramoro_pre}
\end{figure}

In this section, we use the Kuramoto model  to demonstrate the usefulness 
of the dataset on training real roots classification models.
For this purpose, three machine learning methods implemented in Python's scikit-learn library, including K-Nearest Neighbors (KNN), Support Vector Machine (SVM), and Random Forest, are employed.
These models take a given value of parameters as input and predict the number of real solutions 
of the Kuramoto model. 
The dataset for the Kuramoto model contains 10,000 examples and it is split into the training set,  the validation set  and the testing set in the ratio of 8:1:1.
%To simplify the implementation, we directly utilized the corresponding KNeighborsRegressor, SVR, and RandomForestRegressor models from Python's scikit-learn library. 
The experimental results on the testing set are illustrated in Fig.~\ref{fig:kuramoro_pre}. 
For the purpose of visualization, the results have been projected on the parameter space $(w_1,w_4)$. 
From the figure, we observe that the KNN model can classify the number of real roots 
of the parametric Kuramoto model at a very high accuracy (93.3\%). 
Hyperparameter tuning was performed for the three baseline machine learning models on the validation set of the Kuramoto system. The KNN model achieved its best performance with k=2, attaining an accuracy of 93.3\%. The SVM model obtained its highest accuracy of 55.1\%, when using the radial basis function kernel with penalty parameter C=10, where C determines the balance between margin maximization and training error penalization. For Random Forest, the optimal setting was n\_estimators=40, where n\_estimators denotes the number of trees in the ensemble, and this configuration achieved an accuracy of 89.8\%.
% Table~\ref{tab:model_accuracy_comparison} further demonstrates how the hyperparameters 
% for each model are chosen. 

\section{Conclusion and future work}
In this work, we have made a comprehensive search of literatures and repositories on subdivision to collect  nonlinear equations with finitely many real solutions.
We identify and remove duplicate examples combing from different resources, 
call state-of-the-art subdivision solvers to compute real solutions, 
and further verify these results with symbolic solvers. 
As a result, we obtain so far the largest labelled real dataset of zero-dimensional nonlinear equations 
for subdivision. We further select 5 families of nonlinear parametric systems coming 
from different application domains, randomly sample values from the ranges of parameters, 
specialize at these parametric values to generate more than 48K zero-dimensional nonlinear equations, 
and call IbexSolve to compute the real solutions to enhance the dataset. 

We demonstrate that the dataset can  benchmark different solvers
as well as training machine learning models. 
When we develop the dataset, most of the examples have been solved by the state-of-the-art subdivision solver IbexSolve. 
During the development of the dataset, we also identify 
several weakness of the default strategies of IbexSolve,
which creates opportunities in the future to develop more efficient heuristic or learning-based strategies. 

	\section*{Acknowledgment}
	The authors would like to thank J\"urgen Gerhard for helping us getting Maple licence.
%	We would like to thank the anonymous reviewers for their valuable comments and suggestions.

	% 基金信息
	\section*{Funding information}
	This research was supported by the following grants: the National Key Research Project of China under Grant No. 2023YFA1009402, the National Natural Science Foundation of China under Grant No.12301650 and Chongqing Talents Plan Youth Top-Notch Project  under Grant No. 2021000263.
	
	% 作者贡献

% 	\section*{Author contributions}
% 	[Each author's contribution can be specific to the appropriate role, such as "data analysis", "literature review", etc., as described in the standard NISO CRediT \url{https://credit.niso.org/}.]
	
% 	XXX (XXX@XXXX; ORCID: XXXX): Writing - Review (Lead); Methodology (Equal).
	
% 	XXX (XXX@XXXX; ORCID: XXXX): Conceptualization; Formal analysis(Equal); Writing - Original Draft (Lead).
	
% 	XXX (XXX@XXXX; ORCID: XXXX): Methodology (Lead); Formal analysis (Equal), Writing -	original draft (Supporting).

	% 数据可用性声明
	\section*{Data availability statements}
% 	Describe the source and authorization of the data. The data that have been analyzed in this study are available from the author upon request.
	The data that support the findings of this study are openly available in Science Data Bank at \url{https://doi.org/10.57760/sciencedb.34211}. Please cite the source when using the data.

\bibliographystyle{plainnat}
\bibliography{reference}

\end{document}